\documentclass[journal]{IEEEtran}

\usepackage{xcolor,soul,framed} %,caption
\usepackage{minted}

\colorlet{shadecolor}{yellow}
\usepackage[pdftex]{graphicx}
\graphicspath{{../pdf/}{../jpeg/}}
\DeclareGraphicsExtensions{.pdf,.jpeg,.png}

\usepackage{amsmath} % assumes amsmath package installed
\usepackage{amssymb}  % assumes amsmath package installed
\usepackage{amsthm}
\usepackage{array}
\usepackage{mdwmath}
\usepackage{mdwtab}
\usepackage{bm}
\usepackage{eqparbox}
\usepackage{url}
\usepackage{subfigure}
\usepackage{tabularx}
\usepackage{wrapfig}
\usepackage{multirow}
\usepackage{tikz}
\usepackage{graphicx}
\usepackage{subcaption}
\usepackage{booktabs}

\usepackage[justification=centering]{caption}

\usepackage{fancyhdr}
\usepackage{lipsum} % Only needed if you want placeholder text for testing

\usepackage{fancyhdr}
\begin{document}
\bstctlcite{IEEEexample:BSTcontrol}
    \title{COSMO: O-RAN-Based Service Management and Orchestration for Cross-Technology Multi-Tenant Radio Access Networks}
 \author{Miguel Catalan-Cid, Joan Josep Aleixendri, Jorge Pueyo, Pau Tomas, Daniel Camps-Mur 
\thanks{Jorge Pueyo is with UPC, Barcelona, Spain (jorge.pueyo@upc.edu). Pau Tomas is with Satelliot, Barcelona, Spain (pau.tomas@sateliot.com).  The other authors are with i2CAT Foundation, Barcelona, Spain (\{miguel.catalan, daniel.camps, joan.aleixendri\}@i2cat.net). 
Manuscript received Month XX, 20XX; revised Month XX, 20XX.}}

% ====================================================================
\maketitle
\thispagestyle{fancy} % applies top note to the title page

% === ABSTRACT ====================================================================
% =================================================================================
\begin{abstract}
The evolution toward 6G networks envisions a heterogeneous Radio Access Network (RAN) comprising diverse access technologies, such as private 5G, public 4G/5G, and Wi-Fi, managed by multiple stakeholders. %This vision requires novel management frameworks that enable resource sharing across technologies and ensure multi-tenancy support. 
{While considerable research effort has been devoted to O-RAN-based frameworks enabling rApp and xApp implementation and validation, few works provide integrated support for cross-technology RAN orchestration, end-to-end multi-tenancy, and a unified subset of SMO functionalities, including Non-RT RIC components.}
This paper introduces COSMO, a novel RAN Service Management and Orchestration platform designed to support heterogeneous 3GPP (5G NR, LTE) and non-3GPP (Wi-Fi) access networks. {COSMO enables cross-technology multi-tenancy, defined as the capability to allow multiple tenants to dynamically share heterogeneous RAN resources with explicit resource allocation guarantees based on Service Level Agreements (SLAs). This is achieved through management primitives that support flexible and on-demand resource allocation.}
%COSMO enables cross-technology multi-tenancy by employing management primitives that allow dynamic and flexible resource allocation based on Service Level Agreements (SLAs). 
Additionally, the platform includes a cross-technology Non-Real-Time RAN Intelligent Controller (Non-RT RIC) that enables the development of intelligent rApps for closed-loop control and network orchestration. {Beyond its architectural design, COSMO improves resource utilization and operational flexibility through unified orchestration of heterogeneous multi-tenant RAN resources. }Through prototyping and benchmarking, we demonstrate the effectiveness of COSMO in resource allocation, SLA enforcement, and scalability. %We showcase an example of a multi-tenant cross-technology SLA-based RAN slicing application, highlighting the benefits of dynamic resource allocation in heterogeneous networks. 
{In our prototype, the SLA-based rApp reduces SLA violation from approximately 21\% to below 10\% under dynamic traffic conditions in a heterogeneous RAN deployment including 5G, 4G, and Wi-Fi access networks.}
Our results confirm that COSMO offers an efficient solution for managing and orchestrating future multi-tenant cross-technology RAN environments.
\end{abstract}

% === KEYWORDS ====================================================================
% =================================================================================

% \fi
\IEEEpeerreviewmaketitle

% ====================================================================
% ====================================================================
% ====================================================================

% === I. INTRODUCTION =============================================================
% =================================================================================
\section{Introduction}
\label{sec:introduction}
In the 4G era, only a handful of public mobile networks were typically deployed within a country. However, with the allocation of private spectrum in 5G \cite{5gprivatenetworks}, a large number of private 5G networks are expected to appear in the next few years, which will have to coexist with the public 5G networks. Projecting this trend forward, we can expect the future 6G Radio Access Network (RAN) to be composed of many different RANs, built with heterogeneous technologies and managed by different stakeholders. A vision for the future 6G system is that it should devise mechanisms that allow these heterogeneous resources to be composed on-demand to be able to deliver novel end-to-end connectivity services.

The 6G Infrastructure Association (6G-IA) in Europe released a 6G vision white paper in \cite{6giavision}, proposing a 6G system architecture composed of three layers. First, a heterogeneous resource pool, including compute and connectivity resources managed by different stakeholders. Above the heterogeneous resource pool is a unified controllability layer, which abstracts the heterogeneous resource pool into a set of resources that can be consumed by external parties (tenants). Finally, tenants providing end-to-end services consume the resources offered by the controllability layer. A similar view is shared by the Hexa-X-II project in \cite{hexaXd21}, where it identifies end-to-end service aggregation through federation APIs as a key architectural enabler for the future 6G system. Federation concepts are starting to be applied in the edge computing domain, where the GSMA Operator Group defined a reference architecture that allows different Mobile Network Operators (MNOs) to federate their edge computing resources \cite{edgefederation}. In the RAN domain, however, federation and multi-tenancy have not yet gained commercial adoption.

RAN sharing is nowadays a common practice supported by means of the Multi-Operator RAN (MORAN) model, where two MNOs share a baseband device but radiate in separate frequencies, and the Multi-Operator Core Network (MOCN) model, where both baseband and radio are shared across operators \cite{RanSharing}. However, RAN sharing differs from the true multi-tenancy envisioned for 6G in three essential ways. First, sharing agreements are static, instead of the on demand usage and billing required in multi-tenancy. Second, in RAN sharing there is no explicit reservation of RAN resources for the different MNOs, instead the baseband serves connected User Equipments (UEs) equally as if they were served by the same MNO. Third, RAN sharing agreements are limited to two MNOs, where they alternate the role of infrastructure operator depending on the geographical area. 

Instead, RAN multi-tenancy does not limit the number of tenants and explicitly considers the role of a RAN infrastructure operator separated from the tenant MNOs. {In this work, we define RAN multi-tenancy as the capability of a shared and heterogeneous RAN infrastructure to support multiple independent tenants through explicit and dynamically adjustable allocation of radio resources governed by Service Level Agreements (SLAs). In COSMO, this is realized through network chunks and network services, enabling on-demand resource allocation, cross-technology support, and per-tenant performance guarantees.}

3GPP defined 5G network slicing as a concatenation of physical or virtual network functions tailored to a particular communication service. At the RAN level, network slicing is managed by the RAN Network Slice Subnet Management Function (NSSMF), which configures and controls the RAN functions and resources allocated to the different slices. Thus, network slicing can be seen as an enabler for RAN multi-tenancy whereby a RAN infrastructure operator allocates different slices to different tenants. However, existing frameworks are often limited in scope. For instance, native 3GPP network slicing and its management functions are only able to manage 3GPP-defined radios and are therefore not suited to manage heterogeneous RANs \cite{i2slicer}\cite{5gRanSlicing}. Other proposals like \cite{5g-clarity} define models for heterogeneous private networks composed of both 3GPP 5G NR and IEEE 802.11, but do not discuss how these resources can be allocated across different tenants. Furthermore, a formal alignment of such cross-technology management with the O-RAN architecture, a gap identified in \cite{i2slicer}, remains unaddressed.

The O-RAN Alliance considers RAN slicing and sharing as fundamental use cases to be supported by its proposed architecture \cite{oranusecases}. According to the O-RAN Slicing Architecture \cite{oranslicing}, the Service Management and Orchestration (SMO) component may implement Slice Management Functions, for instance to manage the allocation of RAN resources through the O1 interface. The Non-Real-Time RAN Intelligent Controller (Non-RT RIC) also plays a relevant role in these scenarios by enabling the creation of  control-loops through the so-called rApps to dynamically manage the RAN resources according to slice requirements or SLAs. Required actions, which could be AI-driven, may be directly executed through the O1 interface, or indirectly via A1 policies sent to the Near-RT RIC and then executed by the so-called xApps through the E2 interface \cite{oranusecases}. The management of non-cellular radios, like Wi-Fi, is currently out of the scope of the O-RAN architecture and use cases.  

To the best of our knowledge, {no prior work provides an O-RAN–aligned framework to enable cross-technology RAN service federation}. Therefore, in this paper, we introduce COSMO, a novel Cross-technology O-RAN Service Management and Orchestration platform. COSMO presents three main innovations. First, it supports multi-tenancy and {enables} cross-infrastructure federation of RAN { services through its northbound interfaces, allowing external orchestrators to compose end-to-end services across multiple domains}. This is supported by two novel management primitives, referred to as \emph{network chunk} and \emph{network service}, that allow a RAN infrastructure operator to allocate heterogeneous radio resources across different tenants and to associate them with the metro Ethernet connectivity required to reach each tenant’s core network and services. Second, it realizes these abstractions in an O-RAN–aligned architecture that spans the SMO and Non-RT RIC components and can manage
heterogeneous 3GPP and IEEE 802.11 RAN technologies from different vendors. Third, COSMO provides an SDK and a cross-RAN telemetry subsystem that enables the development of rApps, which implement dynamic Non-RT control-loops to interact with the RAN infrastructure.

Beyond these architectural features, the main evaluated contributions in this paper are threefold: 
\begin{itemize}
    \item [i.] We formalize the COSMO system model and its multi-tenancy abstractions (\emph{network chunks and services}), and we present an architecture and implementation that leverage 3GPP's RAN NSSMF and O-RAN's SMO and Non-RT RIC components to integrate heterogeneous 3GPP and IEEE 802.11 RAN technologies through a unified platform.
    \item [ii.] We prototype COSMO and benchmark its performance and scalability during orchestration and telemetry exposure operations.
    \item [iii.] We design a multi-tenancy-enabling rApp that allows COSMO to meet cross-technology SLAs in terms of assigned radio resources, and we demonstrate a prototype of this rApp in a laboratory testbed.
\end{itemize}

The rest of the paper is organized as follows. Section \ref{sec:sota} presents a review of the state of the art relevant to COSMO. Section \ref{sec:arch_design} presents the design of COSMO, while Section \ref{sec:arch_implementation} details its implementation. Section \ref{sec:perf_eval} benchmarks the performance of our COSMO prototype during varying orchestration and telemetry collection workloads. Section \ref{sec:sla_rapp} presents the design and evaluation of the multi-tenancy rApp. Finally, Section \ref{sec:conclusions} summarizes and concludes this paper.

\section{State of the Art}
\label{sec:sota}
{The realization of multi-tenant network infrastructures has been widely studied in Software-Defined Networking (SDN) and programmable transport networks. Prior works enable multiple tenants to share common infrastructure with logical isolation, for instance through OpenFlow-based slicing and multi-provider access network sharing \cite{openfaas}\cite{ sdn_fwa}, while also addressing authentication and authorization mechanisms for controlled access \cite{sdn_sec}.}

{These approaches operate primarily at the transport layer and are constrained by limited resource partitioning capabilities, authorization risks when exposing programmable resources, and contention among tenants. Similar challenges also arise in RAN environments, where resource sharing depends on the capabilities of the underlying radio technologies, vendor implementations, and SLA requirements across heterogeneous RAN systems.}

The adoption of 5G has fueled the development of RAN management tools, providing features such as RAN slicing, RAN sharing or disaggregated RAN. Complementing these features, network intelligence is required to tame the increased complexity of 5G networks. To this end, the O-RAN architecture represents a step forward in network automation through the specification of the Non-RT and Near-RT RICs plus their associated rApps and xApps to create, respectively, non-RT (1 second and longer) and near-RT (10 milliseconds to 1 second) control-loops implementing intelligent Self-Organizing Network (SON) and Radio Resource Management (RRM) mechanisms. In this section we analyze the main features of relevant frameworks from the state-of-the-art, with a focus on orchestration, monitoring and automation capabilities. 

\subsection{Framework Evaluation Criteria}

The following criteria define the main requirements to provide a fully functional cross-technology multi-tenant SMO:
\begin{itemize}
    \item \emph{C1} \emph{Static configuration of 4G/5G resources}: The baseline ability of a platform to configure day-0 parameters of cellular base stations, such as setting up cells, bands, bandwidth, and TDD patterns.
    \item \emph {C2} \emph{Orchestration of RAN services}: Beyond static configuration, this criterion considers the platform's ability to use higher-level abstractions for the on-demand composition of RAN resources into end-to-end services, such as creating a slice for a specific tenant.
    \item \emph {C3} \emph{Telemetry collection}: The capability to monitor and collect key performance indicators and metrics from the RAN nodes. 
    \item \emph {C4} \emph{Multi-tenancy support}: The platform's native ability to support multiple tenants (e.g., MNOs, private verticals) on a shared infrastructure, including RAN and backhaul resources. 
    \item \emph {C5} \emph{Dynamic configuration of 4G/5G}: The capacity to support control loops by applying new configurations dynamically without service interruption, such as updating per-tenant resource allocation. 
    \item \emph {C6} \emph{Automated configuration reconciliation}: The ability to actively monitor the RAN nodes for configuration issues (e.g., due to reboots or manual errors) and automatically reconcile its state back to the desired one.
    \item \emph {C7} \emph{Cross-technology support}: The feature to manage heterogeneous technologies, specifically 3GPP (4G/5G) and non-3GPP (Wi-Fi) radios, from a single, unified management platform.
    \item \emph {C8} \emph{rApp SDK}: The existence of a software development kit or a well-defined API allowing rApps to consume telemetry (C3) and enforce new configurations (C5), enabling Non-RT O-RAN automated control loops. 
\end{itemize}

It is worth noting that these criteria intentionally omit two advanced features: Near-RT control loops and integrated AI/ML support. This omission is deliberate, as these functions are not core to the paper's main focus. 
\begin{itemize}
    \item Regarding Near-RT control, the focus of this paper is on multi-tenancy management and service orchestration (e.g., for MNOs or private verticals), for which Non-RT control loops (i.e., $> 1$ second) provide sufficient dynamicity. Also, as demonstrated in \cite{oran-exp}, the integration of Near-RT RICs across multi-vendor environments still requires substantial software adaptation and version alignment of E2 components, whereas Non-RT RIC integration is comparatively stable. This reinforces the industry’s current pragmatic preference for Non-RT RIC deployments.
    \item Secondly, the application of AI/ML to SMO frameworks is still an evolving field, with proposals ranging from embedding ML models within individual rApps to the development and integration of external AI/ML frameworks \cite{aismo}. 
\end{itemize}

Given this context, we do not consider Near-RT or AI integration as baseline requirements for the platforms under review. Nevertheless, for the sake of a comprehensive comparison, we mention these capabilities if present in the reviewed solutions. Indeed, we have also explored early integrations of COSMO with Near-RT RICs (through the A1 interface) and AI/ML frameworks in recent work \cite{infocom_demo}.

\subsection{State-of-the-art frameworks}

5G-EmPOWER\footnote{https://5g-empower.github.io/} was one of the first open-source RAN Controllers that implemented abstractions to remotely manage RAN resources. Originally focused on Wi-Fi networks \cite{5mpowerwifi}, it evolved to a 5G cross-technology framework by adding 4G support \cite{5gempower5g} and enabling programmability through a Python SDK and the OpenEmpower protocol. This enables control-loop automations like slicing, load balancing and mobility control. However, the project has been discontinued, and does not integrate 5G or O-RAN capabilities. 

FlexRAN was a former open-source project that enabled remote configuration and programmability of 4G base stations through a real-time controller, control protocol, and standardized APIs \cite{flexran}. The project evolved into FlexRIC\footnote{https://gitlab.eurecom.fr/mosaic5g/flexric}, a 5G O-RAN–compliant Near-RT RIC framework developed by the OpenAirInterface (OAI) community. FlexRIC’s architecture comprises an agent library and a server library, where xApps enable rapid development of use-case-specific controllers \cite{flexric}. Compared to other Near-RT RIC implementations, FlexRIC offers a lightweight and resource-efficient environment, minimizing latency and overhead. Through E2 Service Models such as Key Performance Measurement (KPM) and RAN Control (RC), it supports diverse near-real-time control loops, including traffic steering and slicing. However, it does not natively support orchestration, management, configuration reconciliation, or Non-RT RIC (A1/R1) functionalities. The project remains active, with continuous development and updates from OAI contributors.

SD-RAN is a former Open Networking Foundation (ONF) project implementing a 5G O-RAN Near-RT RIC. Recently, it has been integrated within the Aether project of the Linux Foundation\footnote{https://aetherproject.org/sd-ran/}. The platform centers around a micro-ONOS-based Near-RT RIC, structured with micro-services that handle specific roles. These micro-services manage key O-RAN interfaces (E2, A1, O1) and allow xApps to interact with them via APIs. A provided SDK simplifies xApp development, and includes some functional xApps focused on KPM collection, traffic-steering and RAN slicing. It lacks support for orchestration and management, reconciliation, and Non-RT RIC control (A1 and R1 interfaces), although it can be integrated with other Aether and ONOS tools that may provide these functionalities. The SMaRT-5G initiative from Aether also explores the integration of SD-RAN with AI/ML workflows. 

\begin{table*}[t]
\centering
\caption{Comparison of state of the art RAN management frameworks.}
\renewcommand{\arraystretch}{1.5} % More vertical space
\setlength{\tabcolsep}{4pt} % Adjust column separation
\begin{tabular}{|p{1.5cm}|p{1.8cm}|p{1.8cm}|p{1.5cm}|p{2cm}|p{2cm}|p{1.8cm}|p{1.4cm}|p{1.4cm}|}
\hline
\textbf{Framework} & 
\textbf{Static Config}\newline\textbf{(C1)} & 
\textbf{Orchestration}\newline\textbf{(C2)} & 
\textbf{Telemetry}\newline\textbf{(C3)} & 
\textbf{Multi-tenancy}\newline\textbf{(C4)} & 
\textbf{Dynamic Config}\newline\textbf{(C5)} & 
\textbf{Auto Reconcil}\newline\textbf{(C6)} & 
\textbf{Cross-Tech}\newline\textbf{(C7)} & 
\textbf{rApp SDK}\newline\textbf{(C8)} \\
\hline 
\textbf{5G-EmPOWER} & Partially, only 4G & Yes & Yes & Based on projects, only RAN segment & Partially, only 4G & No & Yes & Partially, not O-RAN based \\
\hline
\textbf{FlexRAN,}\newline\textbf{FlexRIC} & No & No & Yes, through E2 & Partially, through E2-based RAN slicing & Yes, through E2 & No & No & Partially, focused on xApps \\
\hline
\textbf{SD-RAN} & Partially, through complementary Aether tools & Partially, through complementary Aether tools & Yes, through E2 & Partially, through E2-based RAN slicing & Yes, through E2 & No & No & Partially, focused on xApps \\
\hline
\textbf{OSC} & Yes, through ONAP & Yes, through ONAP & Yes, through ONAP and E2 & Yes, through ONAP and E2-based slicing & Yes & Yes, through ONAP & No & Yes \\
\hline
\textbf{COSMO} & Yes & Yes & Yes & Yes & Yes & Yes & Yes & Yes \\
\hline
\end{tabular}
\label{tab:ran_platform_comparison}
\end{table*}

The O-RAN Software Community (OSC)\footnote{https://o-ran-sc.org/} provides the reference open-source implementation of the O-RAN architecture, encompassing both Non-RT and Near-RT RICs plus the SMO. The project evolves through iterative releases, each aligned with O-RAN specifications and focused on extending specific functional capabilities. Since its publication, significant attention has been devoted to the Near-RT RIC and the E2 interface, which have been widely adopted in research and experimental platforms\footnote{https://openrangym.com/}\footnote{https://www.openaicellular.org/}\footnote{https://docs.srsran.com/projects/project/en/latest/tutorials/source/near-rt-ric/source/index.html}
 to demonstrate novel xApps using simulated, emulated, and real E2 nodes \cite{openrangym}\cite{coloran}\cite{openaic}. 
 
OSC SMO and Non-RT RIC components reuse several ONAP subsystems, which provide built-in capabilities for Operations, Administration, and Maintenance (OAM), data collection, inventory, and policy management. This integration also facilitates multi-tenancy and aligns the architecture with ONAP’s service orchestration model. The Non-RT RIC incorporates the ongoing development of the R1 interface, enabling Service and Data Management and Exposure (SME/DME), rApp lifecycle management, and A1 policy control. The Information Coordination Service (ICS), developed within OSC, serves as a key data exchange framework underpinning the R1 interface. As detailed in Section~\ref{sec:arch_design}, COSMO leverages ICS to implement the R1 interface. Recent versions have started to explore AI/ML integration, introducing MLOps workflows for model training, distribution, and inference across the Non-RT and Near-RT RICs through the A1 and R1 interfaces, aligning with the broader O-RAN vision for intelligent closed-loop automation.

{Recent work has explored O-RAN platforms leveraging specific OSC Non-RT and Near-RT RIC components to implement advanced control frameworks through Digital Twins \cite{dtoran}. However, this approach mainly focuses on DT-based evaluation of optimization mechanisms implemented by xApps and rApps. In contrast, COSMO focuses on developing a lightweight O-RAN-aligned platform that enables cross-technology RAN orchestration, telemetry exposure, and programmable control loops on real testbed deployments. Digital Twin techniques could be leveraged as a future extension of COSMO to enable large-scale experimentation and evaluation of control strategies prior to their deployment in real scenarios.}

As summarized in Table~\ref{tab:ran_platform_comparison}, most state-of-the-art frameworks primarily targeted the Near-RT RIC domain, with comparatively less development devoted to the SMO and Non-RT RIC layers. Among earlier efforts, solutions such as 5G-EmPOWER adopted a conceptually similar approach to COSMO, including support for Wi-Fi integration; however, these platforms have been discontinued and do not provide native 5G or O-RAN compliance. 

The O-RAN Software Community (OSC) framework formally supports all criteria considered, except for cross-technology support. However, its complexity and large footprint, along with version fragmentation and stability issues reported in recent evaluations~\cite{rimedolabs2024nrt}, limit its adoption as a fully integrated solution for RAN management, orchestration, and programmability. {In addition, the need to integrate multiple subsystems (e.g., ONAP modules), which may be at different stages of maturity, makes deployment and operation non-trivial.} In practice, the OSC solution is predominantly used as a reference and compliance validation platform, with ongoing development efforts mainly concentrated in the Near-RT RIC domain. Therefore, practical and integrated open-source implementations for the SMO and Non-RT RIC layers remain less developed compared to their Near-RT counterparts, highlighting a significant gap that we aim to address \cite{understanding-oran}.

\section{COSMO Design}
\label{sec:arch_design}

\begin{figure*}[t]
  \centering
  \includegraphics[width=1.75\columnwidth]{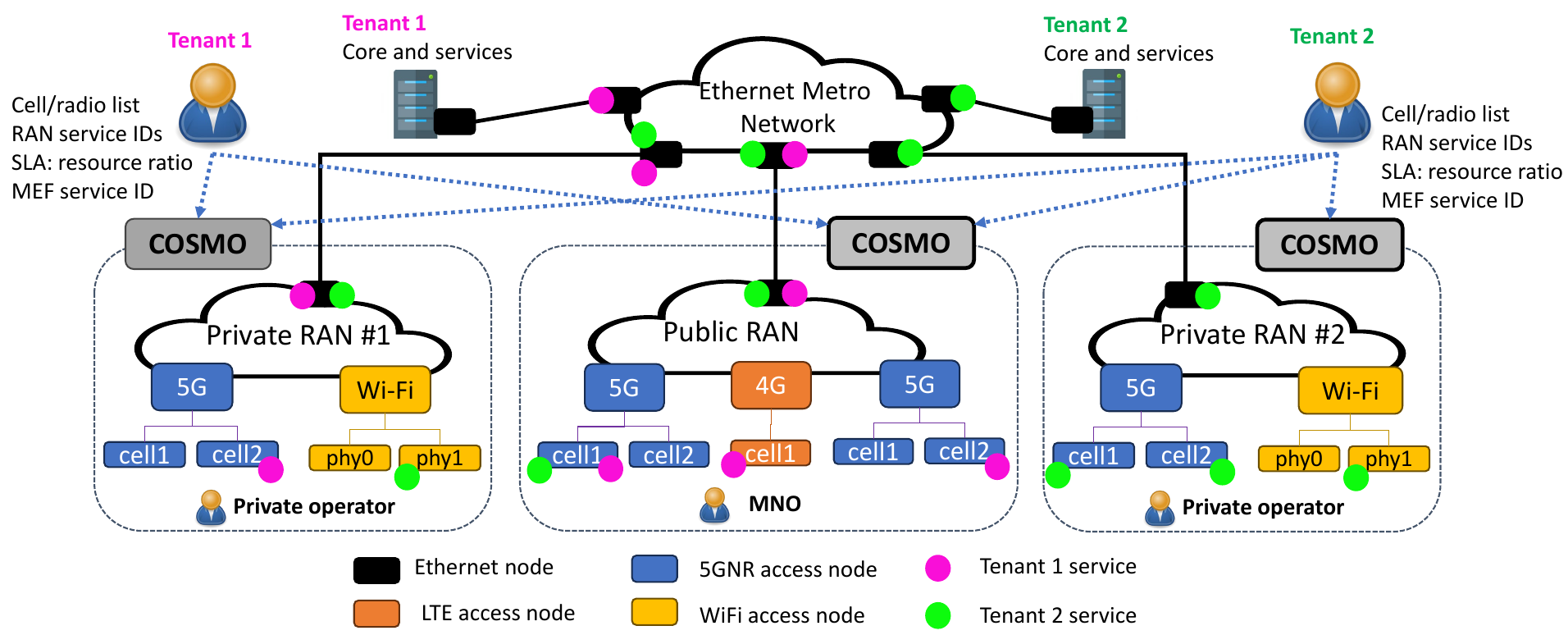}
  \caption{COSMO system model. RAN infrastructure operators deploy COSMO to manage their heterogeneous RAN infrastructure. Tenants request on-demand RAN connectivity services through the COSMO north-bound OpenAPI interface}
  \label{fig:system_model}
\end{figure*}

This section describes the COSMO system model, its multi-tenancy abstractions, and the functional architecture that realizes them. COSMO enables cross-technology multi-tenant RAN services across heterogeneous infrastructures operated by different stakeholders, as illustrated in the COSMO system model in Figure \ref{fig:system_model}. Three RAN infrastructure operators are depicted managing two private RAN networks and a public RAN network. Notice that all these networks are heterogeneous, including LTE, 5G NR and 802.11 radios. Besides the RAN infrastructure operators, Figure \ref{fig:system_model} depicts two tenants, highlighted in pink and green, which consume on-demand RAN connectivity services from the three infrastructure operators. 

The colored circles in Figure \ref{fig:system_model} indicate the physical RAN nodes being allocated to each tenant. Since COSMO focuses on the RAN segment, in the system model indicated in Figure \ref{fig:system_model} tenants are assumed to provide their own core network and services. In the future 6G system though, tenants could consume core network services from other infrastructure operators offering compute resources. The three RAN infrastructures are connected through a metro Ethernet network, and the tenants are able to federate the RAN services offered by each network with their own core network by means of leasing metro Ethernet services from an Ethernet metro operator, e.g. E-line or E-LAN services \cite{mefServices}. 

The key element to enable the on-demand consumption of RAN resources is a COSMO instance that is hosted by each RAN infrastructure operator. COSMO offers a north-bound interface that can be used by the tenants to select RAN resources that will be used to deploy their connectivity services, and to connect these RAN resources to the specific metro Ethernet services required to federate the services consumed from the different RAN infrastructures. The management of Ethernet services across the metro network is however considered to be out of scope for COSMO. Standard SDN controllers can be used for this purpose \cite{SdnWan}.

COSMO models each radio node as an object characterized by its geographical location and a set of radio and backhaul interfaces that can be allocated to one or more tenants. For 3GPP nodes, radio interfaces correspond to cells, which can later be associated with different tenants by leveraging MOCN and/or NSSAI-based slicing. For IEEE 802.11 access points, an interface corresponds to a physical radio (typically one per 2.4/5/6 GHz band), over which multiple tenant-specific services can be advertised using standard virtual-AP mechanisms [25]. Backhaul interfaces (e.g., Ethernet ports) are also part of the node model, capturing the metro Ethernet connectivity that links each radio to the tenant cores and services. 

\begin{figure*}[t]
  \centering
\includegraphics[width=1.75\columnwidth]{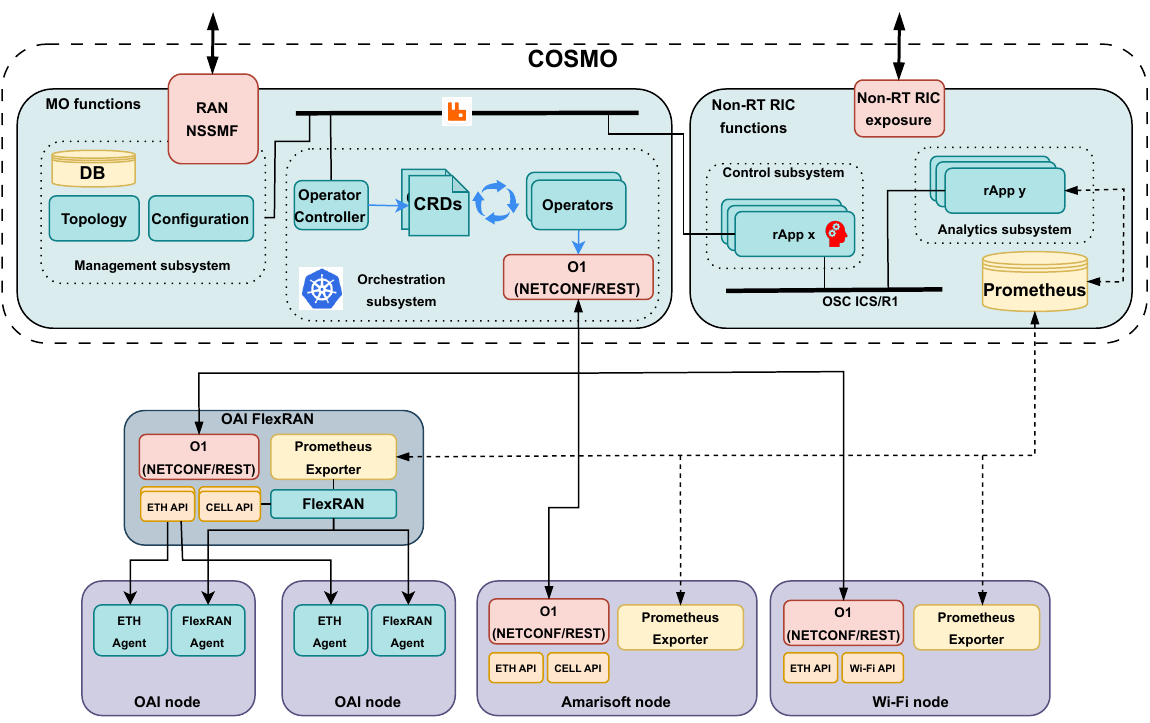}
  \caption{COSMO architecture.}
  \label{fig:COSMO_architecture}
\end{figure*}

\begin{table*}[t]
\centering
\caption{Mapping between O-RAN SMO functionalities and COSMO implementation}
\label{tab:cosmo_smo_mapping}
\begin{tabular}{p{2.4cm} p{7cm} p{7.2cm}}
\toprule
\textbf{O-RAN Functional Block \cite{oransmo}} & \textbf{SMO Description} & \textbf{Implementation in COSMO} \\
\midrule

Service Management and Exposure (SME)
& Provides service registration, discovery, and controlled exposure of SMO services (implemented through the R1 interface in the Non-RT RIC)
& Partially implemented through the Management subsystem API and the messaging bus, allowing rApps to interact with RAN NF OAM \\

\addlinespace
Data Management and Exposure (DME)
& Provides data collection, storage, and exposure services (implemented through the R1 interface in the Non-RT RIC)
& Implemented through the Analytics subsystem, integrating telemetry exporters, Prometheus-based monitoring, and R1/ICS-based information exchange to expose metrics to rApps \\

\addlinespace
RAN NF OAM
& Provides configuration, fault, and performance management of RAN network functions 
& Implemented by the Management and Orchestration subsystems, providing configuration of heterogeneous RAN nodes through NETCONF \\

\addlinespace
 Network Function Orchestrator (NFO)
& Responsible for lifecycle management and orchestration of network functions across the infrastructure 
& Not implemented. Assumed to be handled by external platforms. \\

\addlinespace
FOCOM
& The Federated O-Cloud Orchestration and Management manages lifecycle and orchestration of O-Cloud infrastructure  
& Not implemented. Assumed to be handled by external platforms. \\

\addlinespace
Service and Slice Subnet Orchestration
& Orchestrates services and slice subnet instances across the network infrastructure 
& RAN NSSMF-like functionality implemented via network chunk and network service abstractions, enabling multi-tenant resource allocation across heterogeneous RAN technologies \\

\addlinespace
Service and Slice Subnet Assurance
& Provides monitoring and assurance functions for network services and slice instances 
& Implemented through telemetry monitoring and SLA-driven rApps, which analyze network metrics and dynamically adjust resource allocations \\

\addlinespace
rApp Management
& Provides lifecycle management and execution environment for rApps 
& Implemented through the Kubernetes API in the Non-RT RIC (rApps are managed as pods and services)  \\

\addlinespace
Topology Exposure and Inventory
& Maintains inventory of network resources and exposes topology information to management services 
& Implemented through the Management subsystem database and topology function \\

\addlinespace
Software Package Onboarding
& Supports onboarding and lifecycle management of software packages for network functions and applications 
& Partial/simplified onboarding (K8s-based deployment only) \\

\addlinespace
A1 functions
& Provides A1 interface management between the SMO/Non-RT RIC and Near-RT RIC for policy guidance
& Supported \cite{infocom_demo} but outside the scope of this work \\

\addlinespace
AI/ML Lifecycle Management
& Supports training, onboarding, and deployment of AI/ML models used by rApps 
& Partially supported \cite{infocom_demo} but outside the scope of this work\\

\bottomrule
\end{tabular}
\end{table*}

Figure \ref{fig:system_model} depicts, at high-level, four primitives that tenants use to interact with COSMO: 
\begin{itemize}
    \item The list of cells (3GPP technologies) and physical radios (IEEE 802.11 technologies) that the tenant wants to consume from the RAN infrastructure.
    \item The RAN service identifiers that need to be advertised. These are technology specific and will correspond to the Public Land Mobile Network Identity (PLMN ID) and Single – Network Slice Selection Assistance Information (S-NSSAI) for 3GPP cells, and to the Service Set Identifier (SSID) for IEEE 802.11 radios. 
    \item  The desired share of radio resources, expressed as a percentage of all the available physical resources over the geographical area covered by the requested cells and radios. Notice that the nature of physical resources is different in 3GPP radios, where Physical Resource Blocks (PRBs) are used, than in IEEE 802.11 radios, where airtime is the physical resource being consumed \cite{gadrr}. Hence, COSMO adopts a resource ratio in its SLA definition (cf. Section \ref{sec:sla_rapp}) to be agnostic to the different radio technologies.
    \item The metro Ethernet service identifier, e.g. a VLAN ID, which is used to forward the traffic of the users in each of the RAN infrastructures towards the core and services.
\end{itemize}

To abstract away this heterogeneity in sharing and resource isolation capabilities, COSMO implements two management primitives that enable multi-tenancy across technologies: 
\begin{itemize}
    \item A \emph{network chunk} represents a logical grouping of interfaces that are allocated to a given tenant. A \emph{network chunk}  is composed of cell, radio and Ethernet interfaces, and a given interface can be part of multiple \emph{network chunks}. 
    \item A \emph{network service} is linked to a chunk and defines three main aspects: i) the service identifiers that will be radiated by the access nodes, i.e. PLMNID/SNSSAI for cellular nodes and SSID for Wi-Fi nodes, ii) the fraction of RAN resources allocated to the service for a given technology, if such technology supports sharing of physical resources\footnote{Notice that when a cell or AP belongs to multiple \emph{network chunks} a resource sharing ratio needs to be specified.}, and iii) the VLAN identifiers used by the service to connect to the metro network. To provide all these parameters to COSMO a tenant could make use of an end-to-end slice manager, as described in \cite{i2slicer}.

\end{itemize}

{While COSMO models backhaul interfaces and includes transport-related parameters (e.g., VLAN identifiers) within its service abstraction, the orchestration of metro Ethernet services is intentionally left out of scope. This design choice aligns with the role of the SMO in the O-RAN architecture, where COSMO operates as a RAN NSSMF focused on RAN resource management and orchestration. In contrast, transport network automation is typically handled by dedicated SDN controllers or transport orchestrators (i.e., Transport NSSMFs). COSMO is designed to interoperate with such systems through its northbound interfaces, enabling coordinated end-to-end service provisioning by a higher-level NSMF or slice manager.}

{In this work, federation is understood as the composition of services across multiple independently managed RAN domains through these interfaces, rather than as direct coordination or control between COSMO instances. Accordingly, each infrastructure operator deploys an independent COSMO instance, and multi-domain services are composed by external orchestrators without requiring direct inter-instance coordination. For example, to federate RAN resources across multiple stakeholders, a virtual 3GPP operator would interact with the COSMO instance of each RAN provider. The operator would configure each instance with its own PLMN ID, the VLAN ID corresponding to a shared metro-Ethernet service connecting to the virtual operator’s core network, and the resource allocation ratio agreed upon with each RAN infrastructure provider.}

To realize this vision and turn these abstractions into an operational platform, COSMO adopts a modular layered architecture aligned with the O-RAN SMO components, as shown in  Figure \ref{fig:COSMO_architecture}. {It comprises management, orchestration, analytics, and control subsystems that jointly enable cross-technology RAN orchestration, telemetry-driven analytics, and programmable Non-RT control loops. These subsystems interact through} a lightweight messaging bus (RabbitMQ\footnote{https://www.rabbitmq.com/}), enabling asynchronous communication and scalable deployment. Together, these functions implement full life-cycle management (configuration, monitoring, closed-loop control, and reconciliation) for cross-technology multi-tenant RANs.

{From a security perspective, although COSMO focuses on RAN management and orchestration, authentication and authorization are essential to ensure secure multi-tenant operation. In our design, these functions are assumed to be provided at the platform level via the northbound interface, leveraging standard mechanisms such as API gateways, authentication frameworks (e.g., OAuth2), or token-based access control, as also explored in prior multi-tenant SDN systems \cite{sdn_sec}. These mechanisms enforce tenant isolation by ensuring that each tenant can only access and manage its associated network chunks and services. This isolation applies both to configuration operations and to telemetry exposure, where access to data through the ICS/R1 interface can be restricted based on tenant-specific scopes associated with each service. The integration of COSMO with such security frameworks is left as future work.}

{COSMO is organized into two main functional layers, each comprising multiple subsystems:}

\begin{itemize}
    \item[i.] \emph{MO Functions}: Implement \emph{management} and \emph{orchestration} subsystems responsible for node configuration, \emph{network chunk/service} management, and life-cycle automation (criteria C1, C2, C4-C7). These capabilities are exposed through a REST API leveraging the RAN NSSMF role.
    \item[ii.] \emph{Non-RT RIC Functions}: Enable telemetry, analytics and automated control-loop creation via rApps (criteria C3, C8). The \emph{analytics} subsystem aggregates RAN metrics in non–real-time, while the \emph{control} subsystem exposes telemetry and configuration to rApps implementing control-loops using the COSMO SDK.
\end{itemize}

{In Figure \ref{fig:COSMO_architecture}, the lower part of the architecture corresponds to the RAN nodes and their associated agents, which expose configuration and telemetry interfaces. The MO layer is responsible for configuration and orchestration of these resources, while the Non-RT RIC layer processes telemetry data and enables control-loop execution through rApps. Both layers interact through the messaging bus and shared data services (e.g., ICS/R1), enabling coordination between orchestration and intelligence functions. }

{At a high level, COSMO enables telemetry collection, analytics processing, and control-loop execution across the MO and Non-RT RIC layers, as detailed in \ref{sec:arch_implementation}}. This design ensures clear separation between orchestration and intelligence, while maintaining a common data and messaging layer. {COSMO adopts a logical separation between MO functions and Non-RT RIC functions for deployment modularity and scalability, which is aligned with O-RAN’s service-based approach and decoupled-SMO direction \cite{oransmo}\cite{orandecoupled}.} 

{In summary, COSMO can be understood as a research-oriented subset of the SMO/Non-RT RIC stack, optimized for cross-technology RAN orchestration rather than a full SMO implementation \cite{oransmo}. Specifically, COSMO implements a subset of SMO functionalities focused on RAN orchestration, telemetry exposure, and programmable control loops. In contrast, advanced SMO services such as Network Function Orchestration (NFO) and Federated O-Cloud Orchestration and Management (FOCOM) are not implemented, and are instead delegated to external platforms. A detailed mapping between COSMO and the O-RAN SMO specification is provided in Table \ref{tab:cosmo_smo_mapping}, highlighting the subset of SMO functions implemented in COSMO.}

{The following section describes the implementation of the COSMO architecture, including the integration of heterogeneous RAN technologies with the defined subsystems.}

{
\section{COSMO Implementation}\label{sec:arch_implementation}}

The COSMO prototype is implemented over heterogeneous RAN technologies that are used in the evaluation section. These technologies include Amarisoft 5G NR radios\footnote{https://www.amarisoft.com}, OAI LTE eNBs\footnote{https://openairinterface.org/oai-5g-ran-project/}, and Wi-Fi APs\footnote{https://www.pcengines.ch/apu2.htm} based on the Linux wireless networking stack. The COSMO architecture is designed to be extensible, currently supporting additional LTE, 5G NR, and Wi-Fi vendors, as well as O-RAN emulators, and ready to incorporate future 6G radio technologies as they become available. 

The lower part of Figure \ref{fig:COSMO_architecture} highlights the agents required to integrate with the various COSMO subsystems. To enable remote management and orchestration, both Wi-Fi and Amarisoft nodes embed a NETCONF server. This server exposes two data models: i) the \emph{wifi-api} in the case of Wi-Fi and the \emph{cell-api} in the case of Amarisoft, which allow to configure the respective wireless technologies. In the case of OAI nodes the integration with COSMO is performed through a proxy configuration agent based on FlexRAN \cite{flexran}. In this case, a NETCONF server is collocated with FlexRAN to expose configuration capabilities using the same \emph{cell-api} used for Amarisoft. To manage the configuration of Ethernet metro services COSMO includes an \emph{eth-api}, which is embedded directly in the Amarisoft and Wi-Fi nodes and  allows for remote configuration in the FlexRAN-based OAI nodes. Additionally, all technologies incorporate a telemetry streaming solution that enables Non-RT monitoring and the generation of time-series databases (TSDBs), currently implemented using the open-source Prometheus framework\footnote{https://prometheus.io/}. 

\subsection{Management subsystem}\label{subsec:mngt_section}

\begin{figure}[t]
\begin{subfigure}
  \centering
\includegraphics[width=0.35\columnwidth]{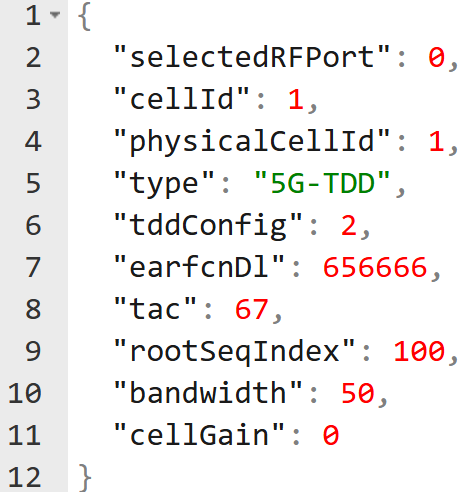}
\end{subfigure}
\begin{subfigure}
  \centering
  \includegraphics[width=0.6\columnwidth]{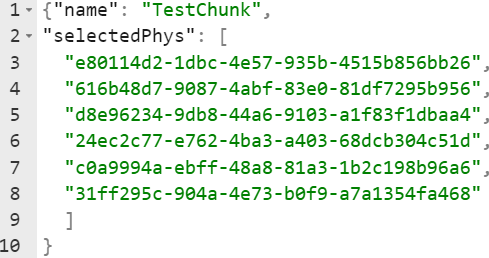}
  \end{subfigure}
    \caption{Example of (a) Amarisoft cell configuration} and (b) POST new network chunk.
    \label{list:post_chunk}
    \label{fig:amarisoft_config}
\end{figure}

The management subsystem provides the APIs and models to enable the remote registration of RAN nodes, the configuration and declaration of \emph{network chunks} (day-0), and the deployment of services over the shared multi-tenant infrastructure (day-1). Persistency is ensured using a PostgreSQL database. The different RAN technologies and vendors are modeled according to their configuration and operation capabilities, which are later translated to NETCONF YANG by the orchestration subsystem. This initial configuration allows to define the static parameters of the RAN nodes, such as the bandwidth, the identifiers (e.g., gnb and cell ids), the transmission power, the channel/bands and operation frequency, or the TDD/FDD configuration, among others. Figure \ref{fig:amarisoft_config}a shows an example of an Amarisoft cell configuration. 

Once registered and configured, the individual interfaces of the nodes can be grouped into \emph{network chunks}. Figure \ref{list:post_chunk}b depicts an example of the POST instruction to create a \emph{network chunk}. Then, multiple connectivity services can be generated by the tenants among the defined \emph{network chunks} which abstract the shared RAN infrastructure.

\begin{figure}[t]
  \centering
  \includegraphics[width=0.7\columnwidth]{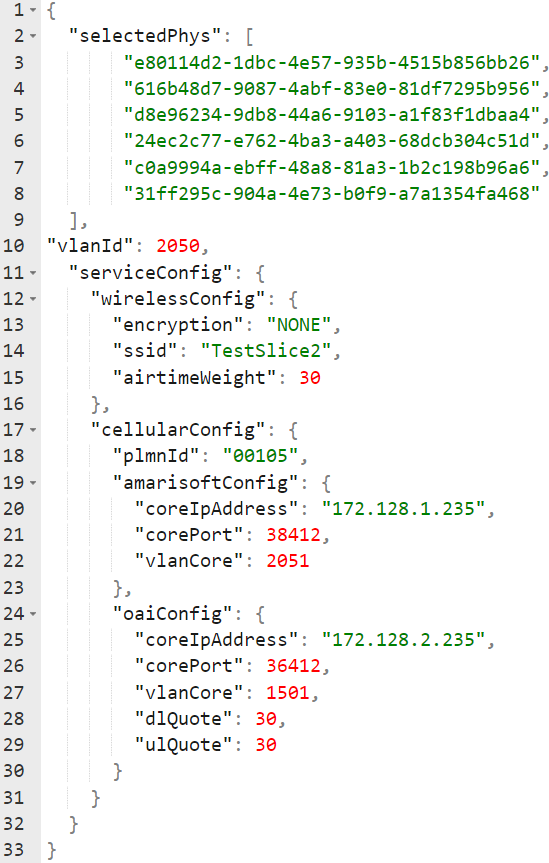}
  \caption{POST new service}
  \label{list:post_service}
\end{figure}

Figure \ref{list:post_service} depicts an example of the POST request to set up a COSMO \emph{service}. As shown in the listing, separate configurations are defined per technology, with LTE OAI and 5G NR Amarisoft including in the cellular configuration the PLMNID to be radiated by the cells, the IP address of the core network they need to connect to, as well as the metro Ethernet VLAN to be used to connect to that core network. The OAI configuration also includes the ratio of PRBs to be allocated to that service. The Wi-Fi configuration includes the SSID to be radiated, the transport VLAN required to connect to the services, as well as the ratio of airtime resources to be applied to that service.  

Note that a challenge to overcome in the quest of enabling multi-tenancy across heterogeneous access technologies is the diverse level of resource isolation capabilities exposed by different technologies. For example, in the case of the three RAN technologies validated in this paper, we encounter the following support for resource isolation:
\begin{itemize}
    \item[i.] LTE-based OAI nodes support sharing through MOCN, which allows to connect a single cell to different core networks owned by different tenants. In addition, through FlexRAN, LTE OAI can slice a single carrier by allocating a fixed number of PRBs to a given tenant. PRB allocations are static, in the sense that if in a Transmission Time Interval (TTI) not all PRBs of a given tenant are used, then these PRBs cannot be reallocated to another tenant. However, a management system is allowed to update the static PRB allocation without impacting ongoing PDU sessions.
    \item[ii.] Linux-based Wi-Fi APs support sharing through virtual APs, which allow to advertise multiple SSIDs over the same Wi-Fi radio \cite{VirtualAps}. In addition, integrating the airtime-based scheduler proposed in \cite{toke}, it is possible to allocate a given portion of airtime to each virtual AP. This local scheduling capability provides the foundation for network-wide, multi-AP slicing policies such as G-ADRR \cite{gadrr}, a concept we extend to a multi-RAT environment in Section \ref{sec:sla_rapp}.  Unlike LTE OAI though, the Wi-Fi airtime scheduler is work-conserving, meaning that unused airtime resources by one virtual AP are automatically redistributed across the other virtual APs.
    \item[iii.] The Amarisoft 5G NR implementation supports sharing through MOCN and 5G SA slicing based on N-SSAI identifiers, but offers no dynamic mechanism to effectively allocate PRBs of a given carrier across different tenants. Therefore, resources among tenants in the same cell are shared according to the default proportional fair scheduling. 
\end{itemize}

{It is important to note that the lack of dynamic PRB allocation in the Amarisoft 5G implementation is a limitation of the specific commercial closed-source stack used in this prototype, rather than of COSMO or 5G systems in general. Other open-source 5G RAN implementations, such as srsRAN or OAI 5G, expose scheduling and slicing controls via E2 interface, enabling PRB allocation through Near-RT RIC control loops. COSMO is designed to accommodate such capabilities through the integration of a Near-RT RIC and E2 Service Models, where Non-RT policies (e.g., via A1 interface) can guide Near-RT control, extending the control loop beyond the scope considered in this work.}

Finally, note that a \emph{network chunk} definition is meant to be fairly static, with existing chunks being modified upon new access nodes being added or removed by the RAN infrastructure operator, or upon changes in the terms of contract between the RAN operator and a given tenant. The service primitive though is meant to be more dynamic, with the control of some service parameters delegated to rApps to enforce some high level SLA. For example, in Section \ref{sec:sla_rapp} we describe the design of an rApp that periodically updates the per-tenant sharing ratio of the services to enforce a high level SLA defined over a given geographic area.

\subsection{Orchestration subsystem}\label{subsec:orchestration}

The objective of the orchestration subsystem is to remotely and dynamically configure the RAN nodes and associated services according to the tenant and rApp demands. As aforementioned, the basis of this system relies on the O1 interface between the SMO and the RAN nodes, which allows using NETCONF and REST APIs. In COSMO, all managed technologies implement a custom YANG model defining their configuration parameters, which are then applied by the agents in each RAN node according to their internal APIs. This approach facilitates extension to new technologies, such as future 6G RANs.

A critical aspect to be addressed by a robust RAN management system is to be able to correct any deviations  introduced due to a variety of reasons such as unexpected reboots or human configuration mistakes. To ensure that the configuration provided to the managed devices is accurate, COSMO  adopts a cloud-native approach and leverages the Kubernetes (K8s) operator framework \cite{cncf_operator_wp}.

A K8s operator consists of a \emph{Custom Resource Definition (CRD)} and an associated \emph{control loop}. A CRD allows the network operator to use the native K8s API to express the desired state of an infrastructure resource in a declarative manner, i.e. using a \emph{YAML} file. The control loop runs inside the K8s cluster and is in charge of periodically checking the current state of the resource, while taking steps to reconcile any deviations detected with respect to the desired state declared in the CRD. In the context of this paper, we define three types of CRDs for each of the supported RAN technologies, namely an \emph{amarisoft-crd}, an \emph{oai-crd}, and a \emph{wifi-crd}. Each of these CRDs is coupled with a dedicated controller that interacts with the NETCONF API hosted by each RAN node (cf. Figure \ref{fig:COSMO_architecture}). The controller runs a reconciliation loop every $T_{controller}$, where it compares the contents of the NETCONF database embedded in each RAN node with the desired state configured in each CRD. If deviations are detected the controller applies the logic in its reconciliation loop to bring the RAN configuration closer to the desired state. 

\begin{figure}[t]
  \centering
\includegraphics[width=0.9\columnwidth]{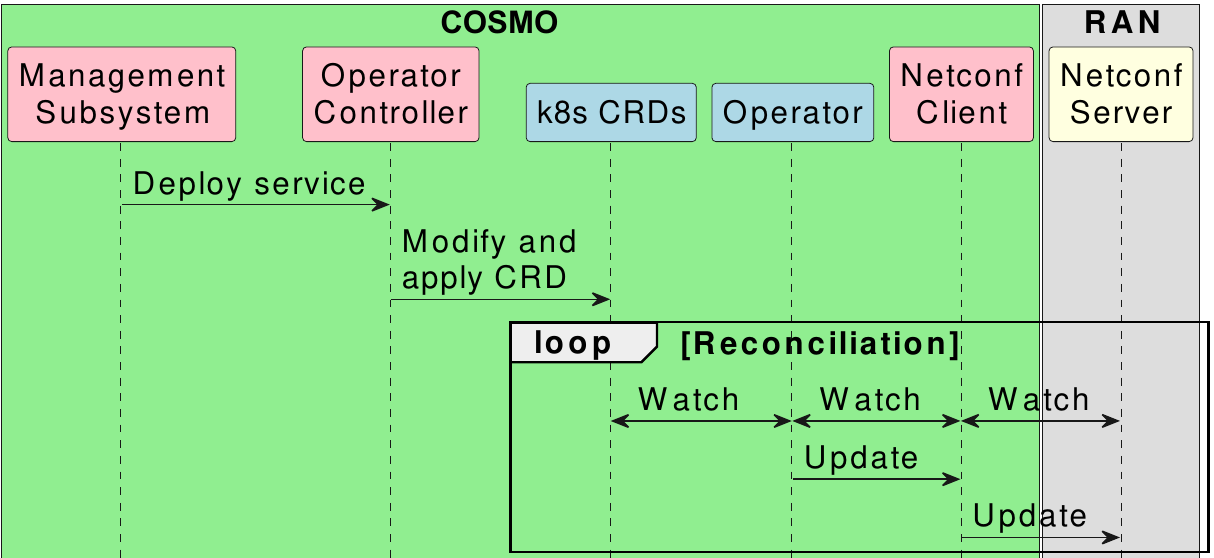}
  \caption{Operator workflow. Reconciliation loop is performed periodically by the operator to ensure infrastructure follows CRD definition.}
  \label{fig:operator-example}
\end{figure}

An illustrative workflow of the orchestration subsystem is provided in Figure \ref{fig:operator-example}, and an experimental evaluation is described in Section \ref{subsec:operator_scal}.

\subsection{Analytics subsystem}\label{subsec:COSMO_telemetry}

The goal of the analytics subsystem is to collect RAN metrics from multiple access nodes in a non–real-time manner (i.e., with reporting periods above one second) and to make these metrics available on demand, either as raw counters or as processed analytics, to control rApps that enforce RAN policies. At the core of the analytics subsystem we have the Information Coordination Service (ICS) component, which is an implementation of the R1 interface by the OSC aimed at providing DME services \cite{isc}. 

\begin{figure}[t]
  \centering
  \begin{subfigure}
    \raggedleft % aligns the image and caption to the left
    \includegraphics[width=0.85\linewidth]{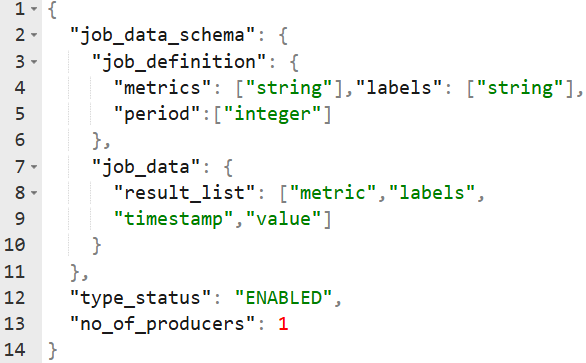}
    \label{fig:ics-example1}
  \end{subfigure}
  \hfill
  \begin{subfigure}
    \raggedleft % aligns the image and caption to the left
    \includegraphics[width=1\linewidth]{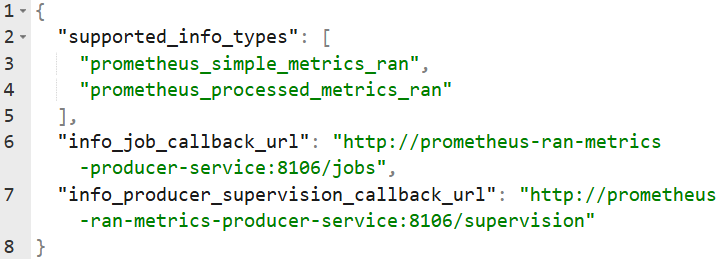}
    \label{fig:ics-example2}
  \end{subfigure}
  \caption{Example of information type and producer rApp for exposing Prometheus metrics using ICS.}
  \label{fig:ics-example}
\end{figure}

The ICS decouples data producers from data consumers, so that they do not need to have any pre-existent knowledge of each other and can be deployed independently. First, Information Types are defined in the ICS to specify the available data. Then, producer rApps capable of exposing available Information types are registered in the ICS (cf. Figure \ref{fig:ics-example}). Finally, consumer rApps interested in any data types create a data subscription through the ICS, which manages the available producers to start serving the data. The exposed data, which may be processed by the producer rApp, can be RAN telemetry, but also external data from other domains (e.g., from application functions or the 5G core) or new data produced by other rApps. Upon receiving and processing the data, the consumer rApp may use it to feed its control-loop logic (e.g., the SLA-based RAN slicing rApp) or to generate new data which is exposed again to other rApps (e.g., rApps doing ML-based anomaly detection, or generating a data set).

\begin{figure}
  \centering
  \includegraphics[width=1\columnwidth]{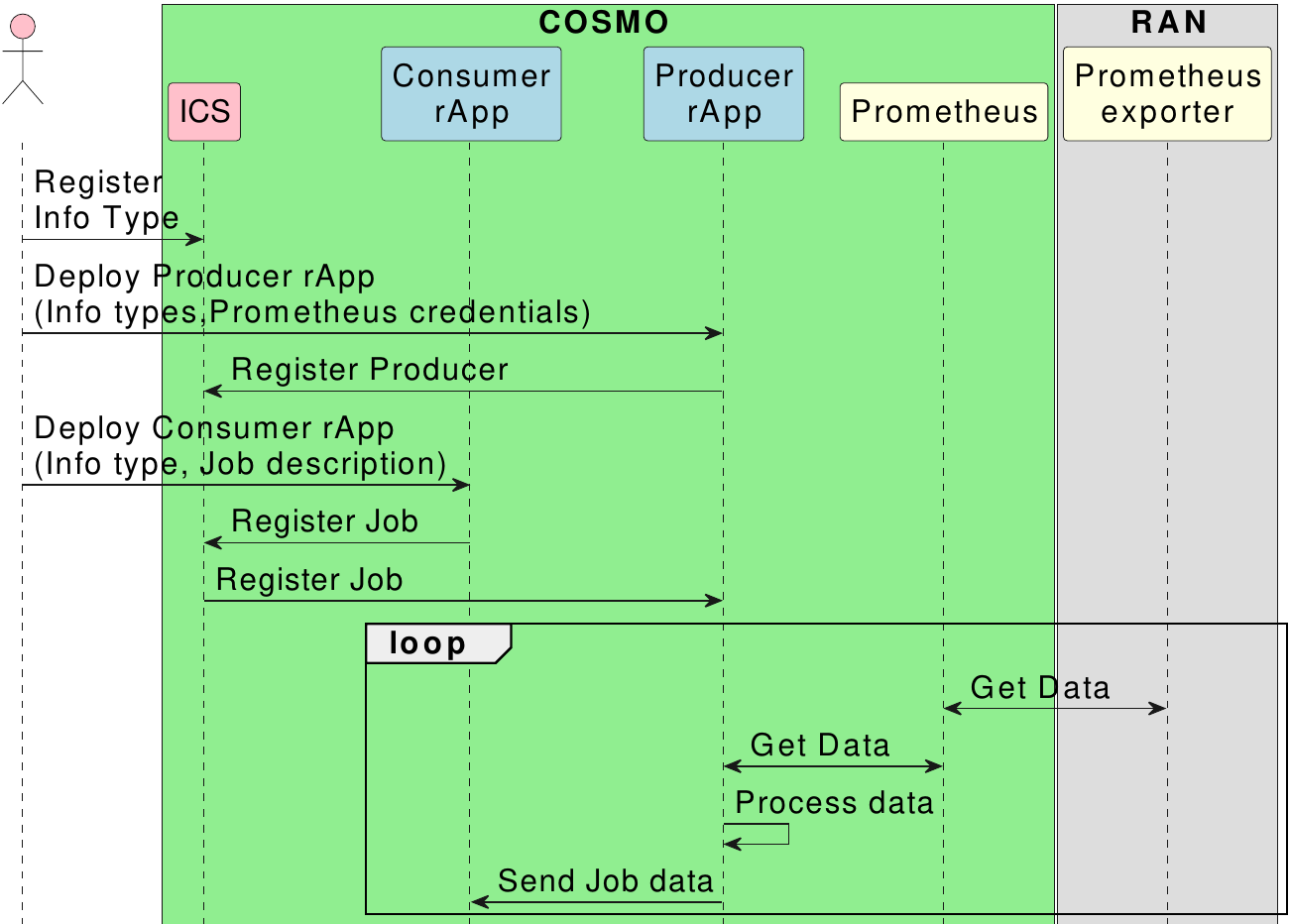}
  \caption{{OSC's ICS/R1-based analytics subsystem: main workflow using Prometheus}}
  \label{fig:ics_workflow}
\end{figure}

By default the COSMO analytics subsystem includes a Prometheus server that stores in a time series database RAN metrics. In the case of the particular implementation reported in this paper, these metrics are scraped from the 4G OAI, 5G Amarisoft and Linux Wi-Fi access nodes (cf. Figure \ref{fig:COSMO_architecture}). Similarly to the NETCONF data model, the Prometheus exporter needs to be customized to each technology. Once deployed, the COSMO analytics subsystem is able to scrape metrics periodically from the different nodes. These metrics are stored in a time series database available to be exposed to the interested rApps. To expose this data to control rApps, the analytics subsystem allows to instantiate one or more analytics producers. For example, we could have a producer for each type of analytic or RAN technology, or a single producer exposing all available analytics. In section \ref{subsec:telemetry_scal} we explore the impact of this choice on scalability.

Figure \ref{fig:ics_workflow} depicts the main workflow involving OSC's ICS, {following the R1 DME producer–consumer paradigm. COSMO leverages the Information Type abstraction defined in ICS to represent cross-technology metrics and processed analytics. This does not introduce any modification to the R1/ICS interfaces or protocols, but rather applies the standard producer–consumer model to heterogeneous data sources.}

In our implementation, the Information Types can be created through the Non-RT RIC or when deploying the specific producers. Information Types can define all or a subset of the metrics stored in a Prometheus server. In the description of the job, the consumer specifies: the Prometheus metric name, the gathering interval, the labels and the Prometheus functions to be applied (e.g., sum, average, max, etc.). Thus, the producer generates, for each registered job, the required query to the Prometheus server and sends the processed data to the consumer. 

The rApp SDK provides flexibility to define new information types and producers that expose processed analytics, as demonstrated in Section~\ref{sec:sla_rapp}, or to publish metrics collected from different telemetry sources, such as those exposed by xApps gathering E2 KPMs. {It is worth noting that COSMO decouples data production and consumption from the underlying telemetry storage through the ICS abstraction. This design enables the integration of alternative telemetry backends (e.g., distributed time-series databases, data buses, or streaming platforms) via dedicated producer rApps, without requiring modifications to the core architecture or consumer logic.}

\subsection{Control subsystem}
\label{subsec:COSMO_sdk}

The control subsystem is composed of the control rApps, which are deployed as pods and services inside the Non-RT RIC Kubernetes cluster leveraging the Kubernetes API, and whose basic life-cycle management (i.e., deploy, delete and status) is controlled by the Non-RT RIC and exposed externally through a REST API. In order to create automated control-loops, the rApps have access to DME services as consumers through the R1/ICS component, as introduced in the previous section, and to the MO management functions through the RabbitMQ.

\begin{figure}[t]
  \centering
  \includegraphics[width=1\columnwidth]{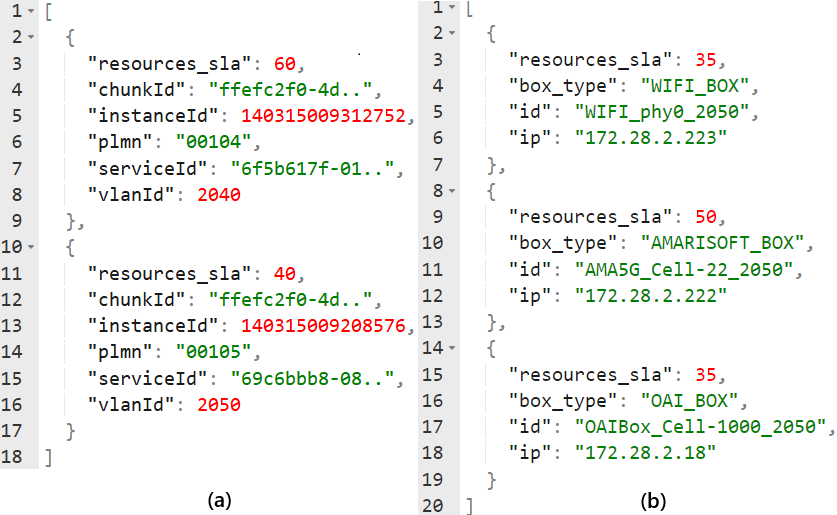}
  \caption{(a) Basic service information and (b) specific information per cell and APs allocated to a service. The id in (b) is used to label the telemetry of this node and service.}
  \label{fig:service-info}
\end{figure}

To illustrate the capabilities of the COSMO rApp SDK, we discuss the capabilities used by the SLA-based RAN slicing rApp described in Section \ref{sec:sla_rapp}: 
\begin{itemize}
    \item[i.] Service topology: Provides the list of cells and APs that are used by a specific \emph{network service}. Each service and each cell or AP has a unique identifier that can be used to link it to the exposed data in the analytics subsystem. Figure \ref{fig:service-info} shows an example of the output of these functions. Interacts with the Topology component of the management subsystem. 
    \item[ii.] Service control: Provides control methods to modify the resource sharing ratio of the whole service or of specific cells or APs, as depicted in Figure \ref{fig:service-info} (i.e., \emph{resource\_sla} field). Interacts with the  Operator Controller of the orchestration subsystem.
    \item[iii.] Data subscription: Managed by the ICS/R1, as presented in Subsection \ref{subsec:COSMO_telemetry}. The control rApps generate jobs or subscriptions according to the definition of the information type, and receive the requested data from the registered producers. 
\end{itemize}

Together, these capabilities enable the creation of automated Non-RT control-loops, which are key to ensure consistent service quality across technologies and tenants, as will be demonstrated in Section \ref{sec:sla_rapp}.

\section{COSMO Scalability Analysis}
\label{sec:perf_eval}

In this section, we benchmark the scalability of our COSMO implementation by studying two possible critical bottlenecks in the required control-loops, namely: i) MO functions enforcing the desired configuration state across RAN nodes, and ii) Non-RT RIC functions managing the exchange of information between producer and consumer rApps.

\begin{figure}[t]
  \centering
  \begin{subfigure}
  \centering
  \includegraphics[width=0.9\columnwidth]{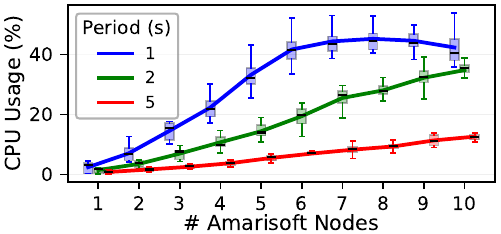}
    \label{fig:scalability_opa}
  \end{subfigure}
    \begin{subfigure}
    \centering
   \includegraphics[width=0.8\columnwidth]{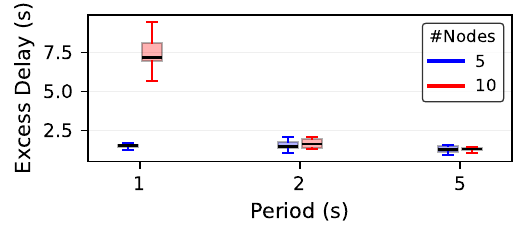}
    \label{fig:scalability_opc}
  \end{subfigure}
    \caption{Operator scalability analysis: (a) CPU usage and (b) excess delay during reconciliation.}
      \label{fig:scalability_op}

\end{figure}

\subsection{MO functions benchmarking}
\label{subsec:operator_scal}
To study the scalability of the MO functions, we focus on the operator framework responsible for configuring and reconciling the RAN infrastructure. We perform several experiments considering different numbers of RAN nodes and reconciliation periods, analyzing their impact on resource consumption and operational delay. To benchmark the operator framework, we focus on the Amarisoft operator, since among the three RAN technologies supported by COSMO, Amarisoft 5G NR requires the most complex configuration and thus the highest computational resources. Note that, for each Amarisoft node considered in the test, the configuration included 10 cells and 2 services per cell, resulting in a large number of operations at each reconciliation period, where the operator inspects the full node configuration to decide if reconciliation is needed.

\begin{figure*}[t]
\centering
\begin{subfigure}
\centering
\includegraphics[width=0.325\linewidth]{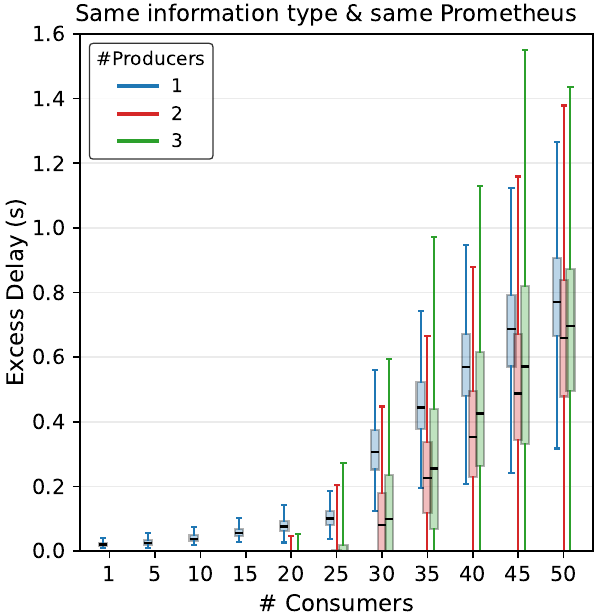}
\end{subfigure}%
\begin{subfigure}
\centering
\includegraphics[width=0.325\linewidth]{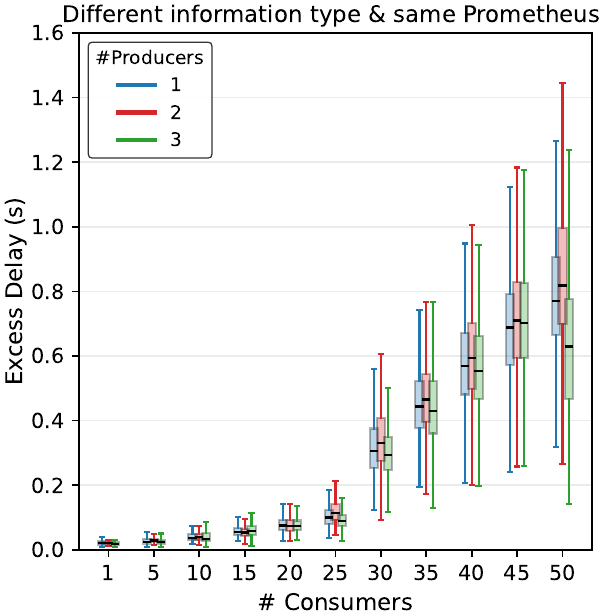}
\end{subfigure}%
\begin{subfigure}
\centering
\includegraphics[width=0.326\linewidth]{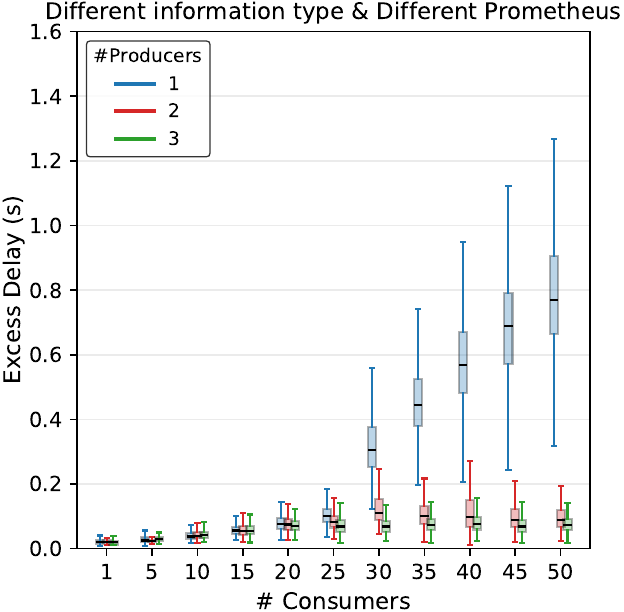}
\end{subfigure}%
  \caption{Experimental results of the telemetry subsystem benchmarking.}
  \label{fig:ics-results}
\end{figure*}

Figure \ref{fig:scalability_op} presents the results of our scalability experiment, where we first monitored CPU as the number of deployed nodes and associated services increased. In this experiment, we used a single pod to run the operator, in order to benchmark its performance without Kubernetes horizontal scaling. Single-pod resource consumption grows significantly with the number of nodes, especially for very short reconciliation periods, due to the high frequency of operations. In the case of a 1-second period, this leads to pod congestion with 7 nodes. This behavior is reflected in the excess delay metric, which represents the difference between the duration of the reconciliation cycle and the configured period. As shown in the figure, only the case with 10 nodes and a 1-second period resulted in an excess that could compromise the reconciliation loop.

Notice that although the results show a clear impact of the number of nodes on the operator’s performance and resource consumption, this effect could be mitigated in realistic scenarios where Kubernetes horizontal scaling could be used, and the required reconciliation periods could be longer, since they focus on semi-static parameters such as cell and service configuration. Dynamic control, instead, will be handled by the RIC functions, as evaluated in the following experiments.

\subsection{Non-RT RIC functions benchmarking}
\label{subsec:telemetry_scal}

To assess the ability of the analytics and control subsystems to exchange data among rApps, we evaluated their performance under different scenarios varying the number of producers, information types, and data sinks. In all cases, each consumer created a job with a required period of 1 second (i.e., the minimum period of a Non-RT control loop), and we computed the excess delay as the difference between the required and the measured periods. New consumers were deployed every 5 minutes.

Figure \ref{fig:ics-results} illustrates the excess delay as a function of the number of producers and the increasing number of consumers. As shown in the left graph, with a single producer, the telemetry subsystem experienced congestion as the number of consumers grew (starting at 25), causing the excess delay to increase linearly. To determine whether the congestion was caused by using a single producer, we repeated the experiment with two and three producers. However, we found that when multiple producers generated the same information type, the ICS replicated each job across all available producers, further increasing congestion in the subsystem. As a result, the excess delay began to increase linearly with a smaller number of consumers. Moreover, this behavior led to some negative excess delays (under 20 consumers) because different producers sent the same data type to the same consumers at slightly different times. In this setup, since new RAN data was generated only once per second, consumers occasionally received repeated, and therefore redundant, data.

In the second scenario, to avoid receiving replicated data, we considered different producers serving distinct information types from a single Prometheus server. As shown in the middle graph of Figure \ref{fig:ics-results}, this configuration did not significantly reduce the excess delay as the number of consumers increased, and the trend remained very similar to that observed with a single producer.  Consequently, we evaluated a third scenario in which each producer served a different information type and obtained data from separate Prometheus servers. In this case, as depicted in the lower graph, the results indicated that this approach alleviated the congestion. This identified the multiple HTTP queries to the API of a single Prometheus server as the root cause of the increasing excess delay.

Based on these results, we can conclude that the implemented Non-RT RIC functions, built upon the ICS/R1 producer–consumer concept, are capable of scaling with the number of rApps. {The observed limitations are primarily associated with the centralized Prometheus backend used in our prototype, rather than with the ICS/R1 architecture itself. Thanks to the decoupling provided by the ICS abstraction, alternative telemetry backends or distributed designs can be adopted, thereby allowing the telemetry infrastructure to be adapted to different scalability and deployment constraints.} For instance, the congestion experienced by Prometheus could potentially be mitigated by increasing the server’s computing resources or by adopting federation approaches such as Thanos\footnote{https://thanos.io/}.

\section{Multi-RAT SLA-based Slicing with COSMO}
\label{sec:sla_rapp}
In this section we illustrate the full capabilities of COSMO through an exemplary use case consisting of an rApp that enforces a RAN slicing SLA over a heterogeneous RAN. We hereafter refer to this rApp as the multi-RAT slicing rApp. 

\subsection{Multi-RAT SLA definition}
\label{subsec:sla-definition}

\begin{figure}[t]
  \centering
  \includegraphics[width=1\columnwidth]{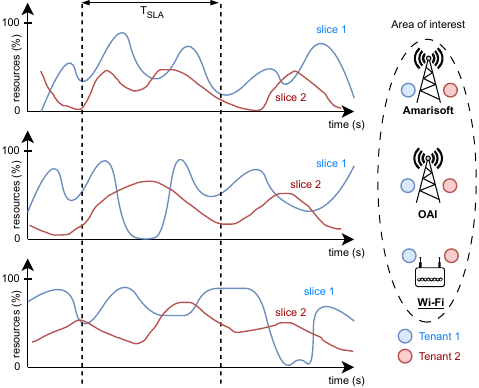}
  \caption{Example of SLA definition}
  \label{fig:rapp_sla}
\end{figure}

Figure \ref{fig:rapp_sla} illustrates a geographic area of interest where a RAN infrastructure provider has deployed three heterogeneous radio access points to provide coverage, namely a 5G Amarisoft node, a 4G OAI node and a Wi-Fi node. Two tenants, shown in red and blue, provide mobile connectivity services to end users over the area of interest using COSMO \emph{network chunks and services}. The relation between the tenants and the RAN infrastructure provider is dictated by an SLA, $0 \leq \rho_{s}^{SLA} \leq 1$, defined over the area of interest, where $s \in \bm{S}= \{ 1, ..., |\bm{S}|\}$ represents the set of \emph{tenants} or \emph{slices}\footnote{We assume that one slice is deployed to serve one tenant and use the two terms interchangeably.} and $\sum_{s \in \bm{S}} \rho_{s}^{SLA} = 1$. 

Note in Figure \ref{fig:rapp_sla} that the load experienced by each slice in each radio node will vary depending on the distribution in time and space of the customer demand for each slice. Let us define $\bm{A} = \{ 1, ..., |\bm{A}|\}$ as the set of radio nodes present over our geographical area of interest. Thus, we can compute the average load observed by slice $s \in \bm{S}$ over the area of interest at a given time as:

\begin{equation}
\rho_{s}^{OBS} (t) = \frac{1}{|A|}\sum_{a \in \bm{A}} \rho^{OBS}_{a,s}(t)
\end{equation}

Let us now define $T^{SLA}$ as a time window over which $\rho_{s}^{SLA}$ needs to be guaranteed. To assess the level of SLA compliance for each tenant over a running window of size $T^{SLA}$ the following operation can be performed:

\begin{equation}
\rho_{s}^{window}(t) = \rho_{s}^{OBS}(t) * \frac{1}{T^{SLA}}w(t)
\end{equation}

, where $w(t)$ is a square signal of duration $T^{SLA}$ and $*$ denotes the convolution operator. This effectively computes a moving average of $\rho_{s}^{OBS}(t)$ over a window of length $T^{SLA}$.

The goal of the multi-RAT slicing rApp described in the next section is to make $\rho_{s}^{window}(t)$ as close as possible to $\rho_{s}^{SLA}$, while accounting for load variations in time and space for each slice, by controlling the resource allocation of each slice in each radio node.

\subsection{Multi-RAT slicing rApp design}
\label{subsec:rapp_design}

The RAN infrastructure operator deploys a COSMO service for each slice, where the service specifies the sharing ratios for each slice in each radio node (cf. Figure \ref{fig:service-info}). We refer to the sharing ratios for tenant $s$ in radio node $a$ as $0 \leq \rho_{a,s}(t+1) \leq 1$. Then, a Multi-RAT slicing rApp is used to periodically update the sharing ratios according to the varying load demands of each slice to meet the SLA. Specifically, the proposed Multi-RAT slicing rApp generalizes the G-ADRR Wi-Fi slicing strategy presented in \cite{gadrr} to be able to operate with the three RAN technologies supported by COSMO. 

The G-ADRR strategy periodically updates $\rho_{a,s}(t+1)$ based on network telemetry obtained in the previous period. Hence, the multi-RAT slicing rApp subscribes to the following telemetry using the COSMO telemetry subsystem:

\begin{itemize}
\item The resource utilization for each slice in each radio node in the last period, i.e. $0 \leq \rho_{a,s}^{OBS} (t) \leq 1$, which corresponds to airtime utilization in the case of Wi-Fi and to cell-level PRB utilization in the case of OAI and Amarisoft.
\item The load offered by each slice in each radio node in the last period, $0 \leq \hat{\lambda}_{a,s}(t) \leq 1$, which measures the average buffer occupancy of the traffic belonging to the slice in the last period, i.e. $\hat{\lambda}_{a,s}(t)=1$ indicates saturation.
\end{itemize}

\begin{figure*}[t]
  \centering
  \includegraphics[width=1.75\columnwidth]{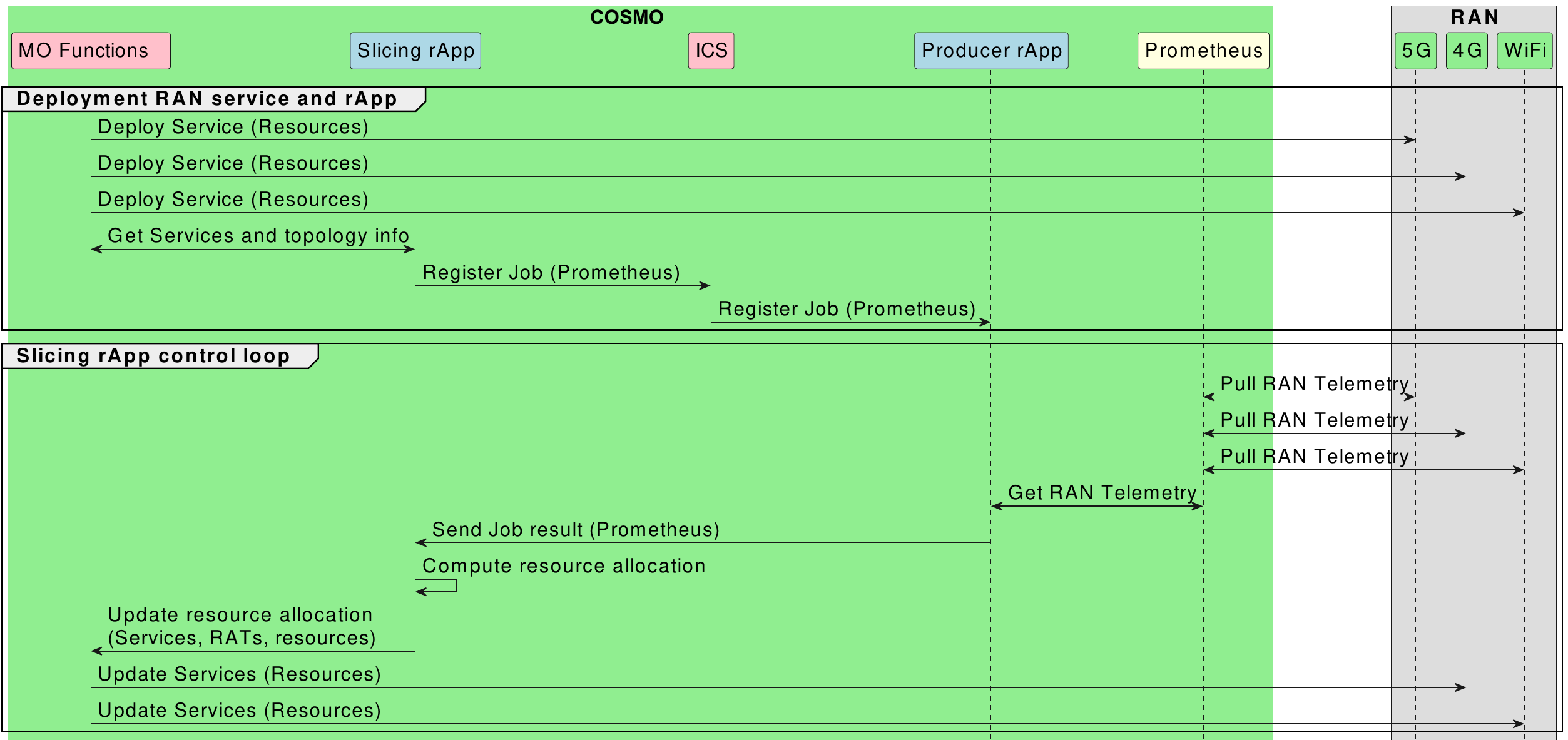}
  \caption{SLA-based slicing rApp workflow}
  \label{fig:rapp_workflow}
\end{figure*}

Based on the previous inputs G-ADRR derives the set of per-slice resource sharing ratios for the next period $\rho_{a,s}(t+1)$ that minimize the cost function described in Equation \ref{eq:gadrr_cost_function}, subject to constraints in Equations \ref{eq:gadrr_constraints_1} and \ref{eq:gadrr_constraints_2}.

\begin{eqnarray}
\label{eq:gadrr_cost_function}
\underset{\rho_{a,s}(t+1)}{\operatorname{argmin}} J_{G-ADRR}(\rho_{a,s}(t+1)) = \\ \nonumber
= J_{1}(\rho_{a,s}(t+1)) + \gamma J_{2}(\rho_{a,s}(t+1)) \\
\label{eq:gadrr_constraints_1}
\sum_{s \in \bm{S}} \rho_{a,s}(t+1) = \rho_{max} \label{eq:J1_c1}  \\
\label{eq:gadrr_constraints_2}
\rho_{min} \leq \rho_{a,s}(t+1) \leq \hat{\lambda}_{a,s}(t+1)\ 
\end{eqnarray}

, where $\hat{\lambda}_{a,s}(t+1)$ is a forecast/estimation of the expected offered load in the next period based on $\hat{\lambda}_{a,s}(t)$, and $0 \leq \rho_{min},\rho_{max} \leq 1$ are two parameters representing respectively a minimum allocation for a slice in a radio node, to avoid users getting locked out during a period, and the maximum capacity that can be allocated in a given radio node. The G-ADRR cost function is composed of two separate cost functions, namely $J_{1}(\rho_{a,s}(t+1))$ and $J_{2}(\rho_{a,s}(t+1))$, where $J_{1}(\rho_{a,s}(t+1))$ is described in Equation \ref{eq:cost_function_J1}:

\begin{equation}
\label{eq:cost_function_J1}
\begin{split}
J_{1}(\rho_{a,s}(t+1))= \sum_{s \in \bm{S}} (\frac{\epsilon}{|\bm{A}|} \sum_{a \in \bm{A}} \rho_{a,s}(t+1) + \\
+ (1-\epsilon)\rho_{s}^{ewma}(t) - \rho^{SLA}_{s})^{2} \\
\end{split}
\end{equation}

This cost function seeks to allocate sharing ratios $\rho_{a,s}(t+1)$ such that an EWMA average of the resource consumption for each slice across the geographical area, i.e. $\rho_{s}^{ewma}(t) = \epsilon\rho_{s}^{OBS}(t) + (1-\epsilon)\rho_{s}^{ewma}(t-1)$, is driven towards the ideal resource ratio, namely $\rho^{SLA}_{s}$. The second cost function $J_{2}(\rho_{a,s}(t+1))$ is described in Equation \ref{eq:cost_function_J2_norm}:

\begin{equation}
\label{eq:cost_function_J2_norm}
\begin{split}
J_{2}(\rho_{a,s}(t+1))=\sum_{s \in \bm{S}} \sum_{a \in \bm{A}} (\frac{\rho_{a,s}(t+1)}{\hat{\lambda}_{a,s}(t+1)} - \\
- \frac{1}{|\bm{A}|}\sum_{i \in \bm{A}} \frac{\rho_{i,s}(t)}{\hat{\lambda}_{i,s}(t+1)})^2\,,
\end{split}
\end{equation}

, where the goal of this cost function is to reward per-slice sharing ratios that have a small deviation across radio nodes, i.e. avoid that a slice has lots of resources in one radio node and very little resources in another one. Note that $J_{2}$ accounts for the fact that the same slice will experience a different load demand in different radio nodes, which is represented by the load estimation $\hat{\lambda}_{a,s}(t+1)$.

As shown in \cite{gadrr} the cost function and constraints in Equations \ref{eq:gadrr_cost_function}, \ref{eq:gadrr_constraints_1} and \ref{eq:gadrr_constraints_2} can be mapped to a Quadratic Programming problem, and thus be computed with any solver in polynomial time.

We adapt the G-ADRR problem definition, which was originally defined only for airtime based allocations in Wi-Fi, to our multi-RAT setup in the following way:
\begin{itemize}
\item [-] \emph{Amarisoft}. Our Amarisoft gNB does not support a local scheduling function that can allocate resources per-slice. This means that the Amarisoft gNB cannot enforce the sharing ratios $\rho_{a,s}(t+1)$ dictated by G-ADRR, instead it will allocate resources to each slice according to its internal MAC scheduler. To account for this shortcoming we adopt the following approach. We measure at each interval the aggregated load that the Amarisoft gNB delivers to each slice, i.e. $\rho_{a,s}^{OBS}(t)$ and force the estimated load in G-ADRR for the next period to match this load, i.e. $\hat{\lambda}_{a,s}(t+1) = \rho_{a,s}^{OBS}(t)$. According to Equations \ref{eq:gadrr_constraints_1} and \ref{eq:gadrr_constraints_2}, this forces G-ADRR to select sharing ratios that match the allocations being performed by the internal MAC scheduler of the gNB. The downside of this approach is that Amarisoft becomes a "passive RAT", and the rApp must compensate for any deviations from the target SLA by adjusting the schedulers of the other technologies.
\item [-] \emph{OAI}. For OAI, we implement a local scheduler that translates the sharing ratios computed by G-ADRR, i.e. $0 \leq \rho_{a,s}(t+1) \leq 1$ into per slice allocated PRBs, represented by \emph{dlQuote} and \emph{ulQuote} fields in Figure \ref{list:post_service}. The PRBs allocated to each slice are updated using FlexRAN every time that G-ADRR takes a new scheduling decision, without requiring any reboot on the OAI nodes. 
\end{itemize}

Figure \ref{fig:rapp_workflow} depicts the detailed operation of our Multi-RAT slicing rApp and its interaction with the COSMO components described in Figure \ref{fig:COSMO_architecture}, which involves two main workflows. First, the $\rho_{a,s}(t+1)$ sharing ratios in each radio node are configured by the rApp through the RAN operators, using the \emph{service} abstraction introduced in Section \ref{subsec:mngt_section}. Second, all RAN devices implement a Prometheus Exporter, enabling per-technology producer rApps in the telemetry subsystem (cf. Section \ref{subsec:COSMO_telemetry}) to gather the RAN metrics required by G-ADRR, such as airtime utilization in the case of Wi-Fi, or cell-level PRB utilization in the case of OAI and Amarisoft. Finally, these two functionalities, RAN monitoring and RAN control, are exposed to the multi-RAT slicing rApp using the SDK subsystem described in Section \ref{subsec:COSMO_sdk}.

\subsection {Testbed description}

\begin{table}[t]
\centering
\caption{Testbed configuration and hardware overview}
\label{tab:testbed}
\centering
\renewcommand{\arraystretch}{1.2}
\footnotesize
\begin{tabular}{|p{8.2cm}|}
\hline
\textbf{Amarisoft (5G)} \\ \hline
\textbf{HW details:} Amarisoft Callbox Advanced + Amarisoft SDRs \\
\textbf{SW details:} Amarisoft FW: 2023-06-10; 5G Core: Open5Gs v2.6.4;\\Custom Prometheus Exporter \\
\textbf{Main RAT configuration:} N77, 4.05 GHz, 50 MHz, 30 kHz, TDD \\
\textbf{UEs:} 2× Laptop + Quectel RM500QL \\ \hline

\textbf{OAI (4G)} \\ \hline
\textbf{HW details:} Intel NUC10i7FNH + Ettus B210 \\
\textbf{SW details:} OAI RAN: 2022.w42; 5G Core: Open5Gs v2.6.4;\\Custom Prometheus Exporter; FlexRAN v2.4 + Custom Scheduler \\
\textbf{Main RAT configuration:} N7, 2.685 GHz, 20 MHz, FDD \\
\textbf{UEs:} 2× Raspberry Pi 4 + SIMCOM 8200EA \\ \hline

\textbf{SBC (Wi-Fi)} \\ \hline
\textbf{HW details:} Pc Engines APU2 apu4d2 + Atheros WLE200NX Wi-Fi \\
\textbf{SW details:} Hostapd v2.7; Custom Local Scheduler;\\Hostapd Prometheus Exporter \\
\textbf{Main RAT configuration:} Channel 149, 20 MHz, 802.11n \\
\textbf{UEs:} 2× Smartphone \\ \hline
\end{tabular}

\end{table}

\begin{figure}[t]
\centering
\includegraphics[width=0.8\columnwidth]{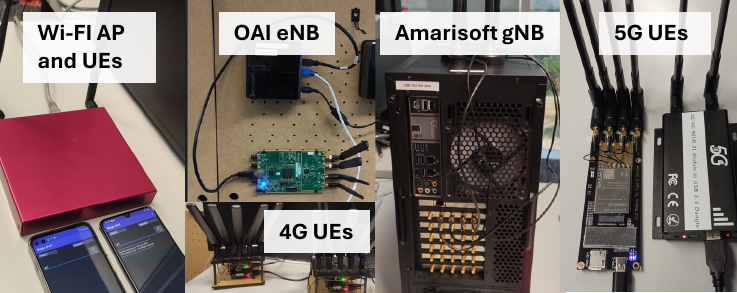}
\caption{Testbed hardware}
\label{fig:testbed_hw}
\end{figure}

\label{subsec:rapp_testbed}
The experimental setup involved the deployment of two RAN services, one per tenant, within a multi-RAT environment, incorporating Wi-Fi, 4G, and 5G technologies. Each tenant in each RAT had one active UE, i.e. six UEs in total were present in the testbed. Table \ref{tab:testbed} summarizes the main characteristics and configuration parameters of the testbed, whose elements are depicted in Figure \ref{fig:testbed_hw}.

{The experimental procedure follows the workflow depicted in Figure \ref{fig:rapp_workflow}. First, after the RAN nodes are remotely configured, the RAN services corresponding to each tenant are deployed across the different RATs using the COSMO orchestration functions. Then, the slicing rApp is instantiated with the service identifiers and requested SLAs, and discovers service and node information through the MO functions. Based on this information, it registers the required monitoring jobs through the ICS/R1 interface. During operation, telemetry data from the RAN nodes, filtered according to the service and node identifiers, is periodically collected and processed by the analytics subsystem. Based on the observed workload and the requested SLA, the rApp computes resource allocation decisions using the proposed algorithm, which are enforced through the management and orchestration subsystems by updating the configuration of the RAN nodes. This process is repeated periodically, forming a closed control loop during the experiment.}

The traffic load in the RAN was generated in the downlink direction, and it was varied according to the requirements of the evaluated scenarios. We used UDP Iperf\footnote{https://iperf.fr/} traffic to avoid any impact of TCP congestion control on the evaluation. In the case of the Wi-Fi and 4G RATs, each of the tenants tried to obtain as many resources as possible (i.e., saturating the channel), forcing the local scheduler to share resources according to the slicing rApp control. In the case of 5G, since the default scheduler is based on proportional fairness, we ensured that the generated traffic load required an amount of resources matching the resources allocated per tenant.

Key Performance Indicators (KPIs) such as throughput and consumed resources per RAT and tenant were captured with Grafana by means of Prometheus Exporters, and exported to CSV files to analyze them. T\textsubscript{SLA} was configured to 100 seconds, and the period of the rApp, T\textsubscript{loop}, was set to 10 seconds (i.e., EWMA with alpha equal to 0.1). The minimum resource allocation per service in a RAN node was configured to $\rho_{min}=4\%$.

\subsection{Evaluation}
\label{subsec:rapp_evaluation}
The evaluation consisted of three different test cases which analyze the performance of the SLA-based slicing rApp under different conditions: (i) Fixed offered load, (ii) dynamic offered load in the 5G RAT, and (iii) dynamic offered load in the Wi-Fi RAT. 

\subsubsection{\textbf{Fixed offered load}}
The aim of the first test case was to validate the rApp operation in cases without load variations. We analyzed the decisions made by the rApp according to different SLA combinations and how they were applied to the RAN by the schedulers. For all the combinations, we compared the results with a case where a static resource allocation per tenant and RAT was enforced during the service deployment, independently of the global area-based SLA defined in (1). Each test lasted around 3 minutes, which was enough to obtain a stable SLA.

\begin{figure}[t]
  \centering
  \includegraphics[width=1\columnwidth]{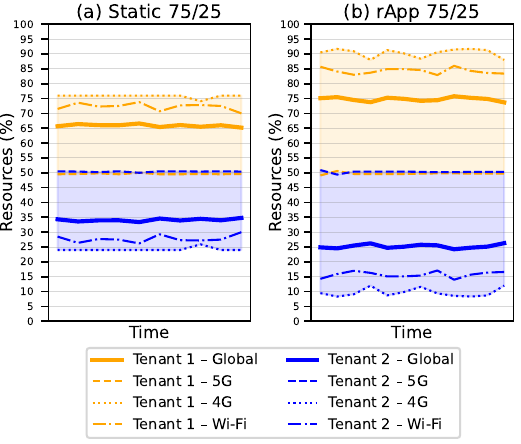}
  \caption{Fixed load, SLA 75\%-25\%: (a) Static and (b) rApp allocation }
  \label{fig:fixed_load_rapp}
\end{figure}

Figure \ref{fig:fixed_load_rapp} illustrates the results of a fixed load scenario targeting SLAs of 75\% (Tenant 1) and 25\% (Tenant 2) using a static and an SLA-based slicing rApp allocation. The figure depicts the variation in resource consumption across the two tenants and the three different radio technologies, which was reported by their respective RAN exporters, and the global resource allocation for each tenant, which was calculated as the average across the three technologies, as defined in (1). Note that in both cases the 5G resource allocation could not be dynamically adjusted and remained at 50\% for each service. As shown in Figure \ref{fig:fixed_load_rapp}a, this caused the static allocation case, fixed during service deployment to 75\%-25\% in all technologies, to violate the global SLA of the tenants by leading to a global resource sharing of 66\%-34\%. On the other hand, as shown in  Figure \ref{fig:fixed_load_rapp}b, the rApp was able to compensate the fixed allocation in 5G by adjusting the resources assigned to the tenants in the 4G and Wi-Fi RATs, obtaining the required global SLAs. 

\begin{figure}
  \centering
  \includegraphics[width=0.8\columnwidth]{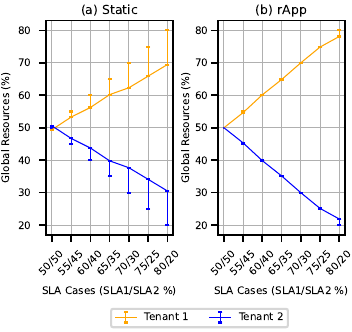}
  \caption{Fixed load, variable SLA: (a) Static and (b) rApp allocation }
  \label{fig:fixed_variable_sla}
\end{figure}

Figure \ref{fig:fixed_variable_sla} summarizes the global SLA results for all the static scenarios tested, where the error bars highlight the SLA violation. Note that, due to the fixed allocation in the 5G RAT, an SLA relationship of 80\%-20\% was the maximum unbalance that could be reached without leading to SLA violations even when considering an optimal allocation: e.g., 85\%-15\% would require to allocate more than the 100\% of resources to tenant 1 in Wi-Fi and 4G RATs. Therefore, we set this limit in this comparison. As expected, the slicing rApp outperformed the static allocation by achieving in almost all the cases the required SLAs and preventing SLA violations up to the 10\%. 

The obtained results also highlight two relevant aspects of the slicing rApp implementation:
\begin{itemize}
     \item[i.] When possible, the algorithm tries to balance the resource allocation across all the available RAN nodes, trying to reach similar tenant allocations among all of them \cite{gadrr}. However, since we had different local scheduling methods for 4G and Wi-Fi, in some cases reaching the optimal balance was not possible. In general, allocation in Wi-Fi offered a higher granularity (airtime percentage), which was used by the algorithm to compensate the less flexible allocation of 4G (number of PRBs).
     \item [ii.] The Wi-Fi RAT was not able to achieve the required resource allocation of approximately 95\%-5\% due to limitations in its local scheduler implementation. This led to a small SLA violation in the 80\%-20\% case. 
\end{itemize}

\subsubsection{\textbf{Dynamic 5G load}}

\begin{figure} [t]
  \centering
  \includegraphics[width=1\columnwidth]{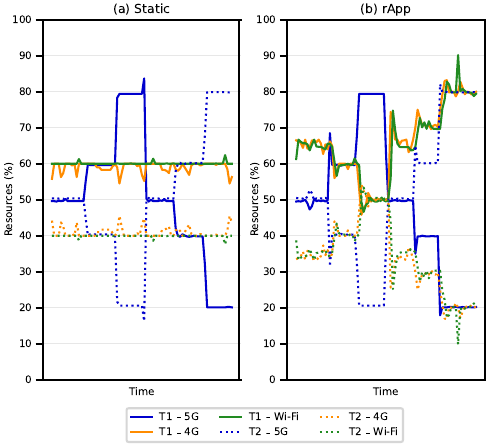}
  \caption{Dynamic 5G load: (a) Static and (b) rApp allocation per technology}
  \label{fig:dynamic1-load}
\end{figure}

In this second scenario, we analyzed the performance of the rApp when dealing with uncontrolled traffic variations in the 5G RAT. Concretely, we considered a fixed SLA of 60\%-40\% among tenant 1 and tenant 2, and each 2-3 minutes we modified the offered load of the tenants in the 5G RAT as follows: 50\%-50\%, 60\%-40\%, 80\%-20\%, 40\%-60\% and 20\%-80\%. These variations are illustrated in the 5G resources graph of Figure \ref{fig:dynamic1-load}.

\begin{figure}[t]
  \centering
  \includegraphics[width=1\columnwidth]{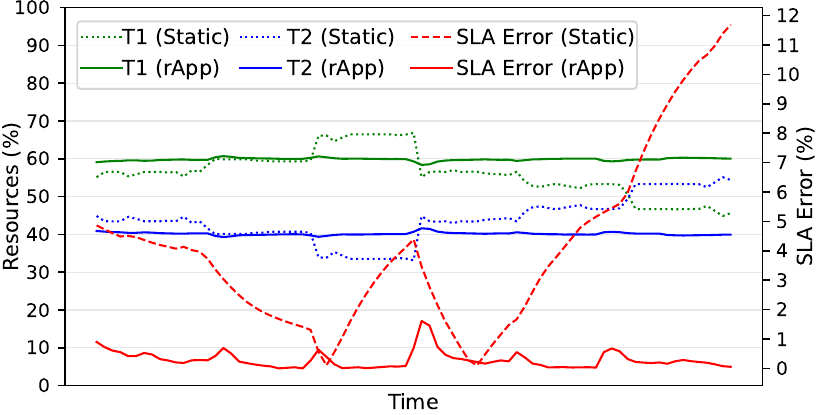}
  \caption{Dynamic 5G load: (a) Static and (b) rApp global allocation and error} 
  \label{fig:dynamic1-load2}
\end{figure}

As depicted in Figure \ref{fig:dynamic1-load}a, the static allocation (i.e. fixed allocation to 60\%-40\% in all the RATs), led to an ideal resource sharing in each single RAT, but to global resource allocation that did not meet the SLA of the tenants. Figure~\ref{fig:dynamic1-load2}a shows the SLA error (secondary Y-axis), reaching up to 12\% in each tenant in the final configuration where tenant 2 obtained globally more resources than tenant 1. The SLA error in Figure\ref{fig:dynamic1-load2} is computed as follows, considering the relationship between T\textsubscript{loop} and T\textsubscript{SLA}:
\begin{equation}
    SLA\textsubscript{error}(t) = \mid SLA\textsubscript{required}-SLA(t) \mid
\end{equation}
\begin{equation}
    SLA(t) = 0.1*SLA\textsubscript{measured}(t)+0.9*SLA(t-T\textsubscript{loop})
\end{equation}

In the case of the SLA-based slicing rApp, as shown in Figure \ref{fig:dynamic1-load}b, the resource allocations of Wi-Fi and 4G RATs were adapted to the traffic variations in the 5G RAT. As shown in Figure \ref{fig:dynamic1-load2}, the maximum SLA error was less than 2\%, and was caused due to the reaction time needed to export the 5G load to Prometheus, expose it through the producer rApp, and apply the new resource allocation. Also, note that punctual SLA errors, caused by traffic or wireless medium dynamics, will be compensated by the rApp algorithm in the next T\textsubscript{loop} periods, thus stabilizing to the required SLA during the time window T\textsubscript{SLA}. Additionally, lower T\textsubscript{loop} periods could be applied in case a faster optimization is needed. 

\subsubsection{\textbf{Dynamic Wi-Fi load}}
The third scenario evaluated the SLA-based slicing rApp behavior in a corner case where it was unable to meet the SLA for a period of time. This was caused by stopping the traffic of the first tenant of the Wi-Fi RAT, which, combined with the uncontrolled scheduler of 5G RAT, made it impossible to reach the global SLA of 60\%-40\%. The test lasted for 12 minutes. 

\begin{figure} [t]
  \centering
  \includegraphics[width=1\columnwidth]{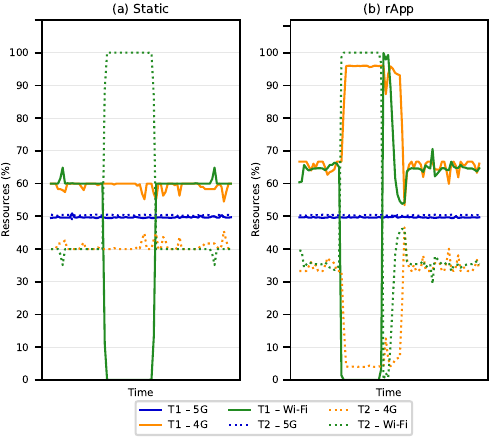}
  \caption{Dynamic Wi-Fi load: (a) Static and (b) rApp allocation per technology}
  \label{fig:dynamic3-load}
\end{figure}

\begin{figure} [t]
  \centering
  \includegraphics[width=1\columnwidth]{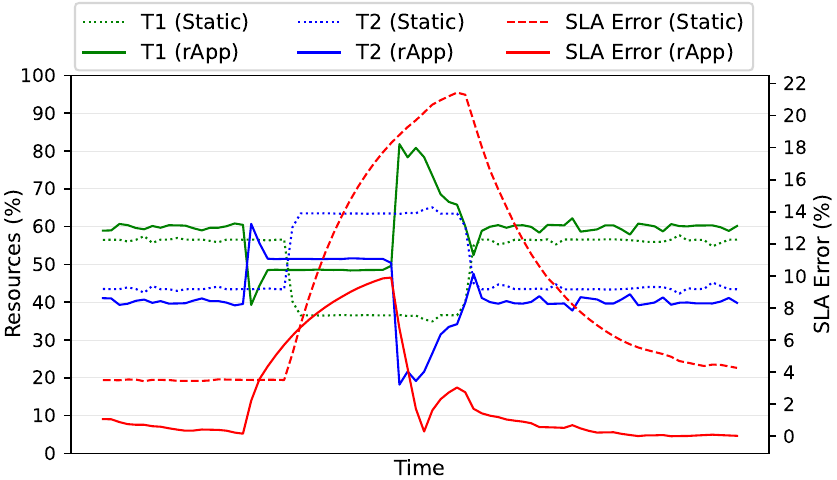}
  \caption{Dynamic Wi-Fi load: (a) Static and (b) rApp global allocation and error} 
  \label{fig:dynamic4-load}
\end{figure}

As shown in Figures \ref{fig:dynamic3-load} and \ref{fig:dynamic4-load}, during this period, the rApp maximized the allocated resources to the first tenant in the 4G RAT; only a $\rho_{min}=4\%$ was allocated to the second tenant. Thus, the rApp maintained the SLA violation below 10\%, instead of the 21\% that was reached by the static allocation. Once the load of tenant 1 in the Wi-Fi RAT resumed, the rApp attempted to compensate for the SLA violation by allocating additional resources to this tenant, reaching a global resource allocation of 80\%-20\% during some T\textsubscript{loop} periods. This caused the SLA error, which is based on Equation (8), to decrease rapidly compared to the static case. Eventually, the resource allocation across all tenants and RATs stabilized to the values prior to the violation.

Overall, these experiments show that the multi-RAT slicing rApp can compensate for uncontrolled RAN nodes and spatio-temporal load variations, enforcing the cross-technology SLA across multi-tenant infrastructures. Combined with the scalability results in Section \ref{sec:perf_eval}, this practical use case validates the COSMO framework end-to-end, illustrating how its management abstractions (C1–C4), robust orchestration (C6), and Non-RT RIC capabilities (C3, C5, C8) enable intelligent and automated control over a heterogeneous RAN.

\section{Conclusions}
\label{sec:conclusions}
In this paper, we presented COSMO, an innovative solution for managing heterogeneous RAN infrastructures composed of 3GPP and non-3GPP technologies. The key contributions of COSMO are its ability to support multi-tenancy across diverse RAN technologies, its comprehensive orchestration capabilities, and its  automated Non-RT closed-loop control through intelligent rApps. Through experimental validation, we demonstrated the platform's scalability, performance, and its ability to enforce dynamic SLA-based RAN slicing in a multi-tenant environment. The use of advanced management abstractions, like network chunks and services, enables precise control over resource allocation while abstracting the complexity of heterogeneous technologies. 

This work takes a step toward cross-technology multi-tenant RAN management and orchestration for 6G. Future work includes the integration of COSMO with Near-RT control to enable 5G O-RAN-based dynamic control, the exposure of O-RAN analytics and policies to external components such as CAMARA APIs, and the integration with AI/ML workflows to enable AI-for-RAN scenarios. 

\section{Acknowledgments}
\label{sec:acks}
This work was supported by the European Union through the SUNRISE-6G project (Grant Agreement No. 101139257). It also received support from the COALESCE-6G project (PID2024-163028OB-I00), funded by MICIU/AEI/10.13039/501100011033/FEDER, EU, and by the Spanish Ministry of Economic Affairs and Digital Transformation and the European Union – NextGenerationEU, under the Recovery, Transformation and Resilience Plan (PRTR) through the UNICO I+D 5G 2021 call (TSI-063000-2021-15 – 6GSMART-EZ). The authors further acknowledge the support of the CERCA Programme of the Generalitat de Catalunya.

\bibliographystyle{IEEEtran}
\bibliography{Bibliography}

\begin{IEEEbiography}[{\includegraphics[width=1in,height=1.25in,clip,keepaspectratio]{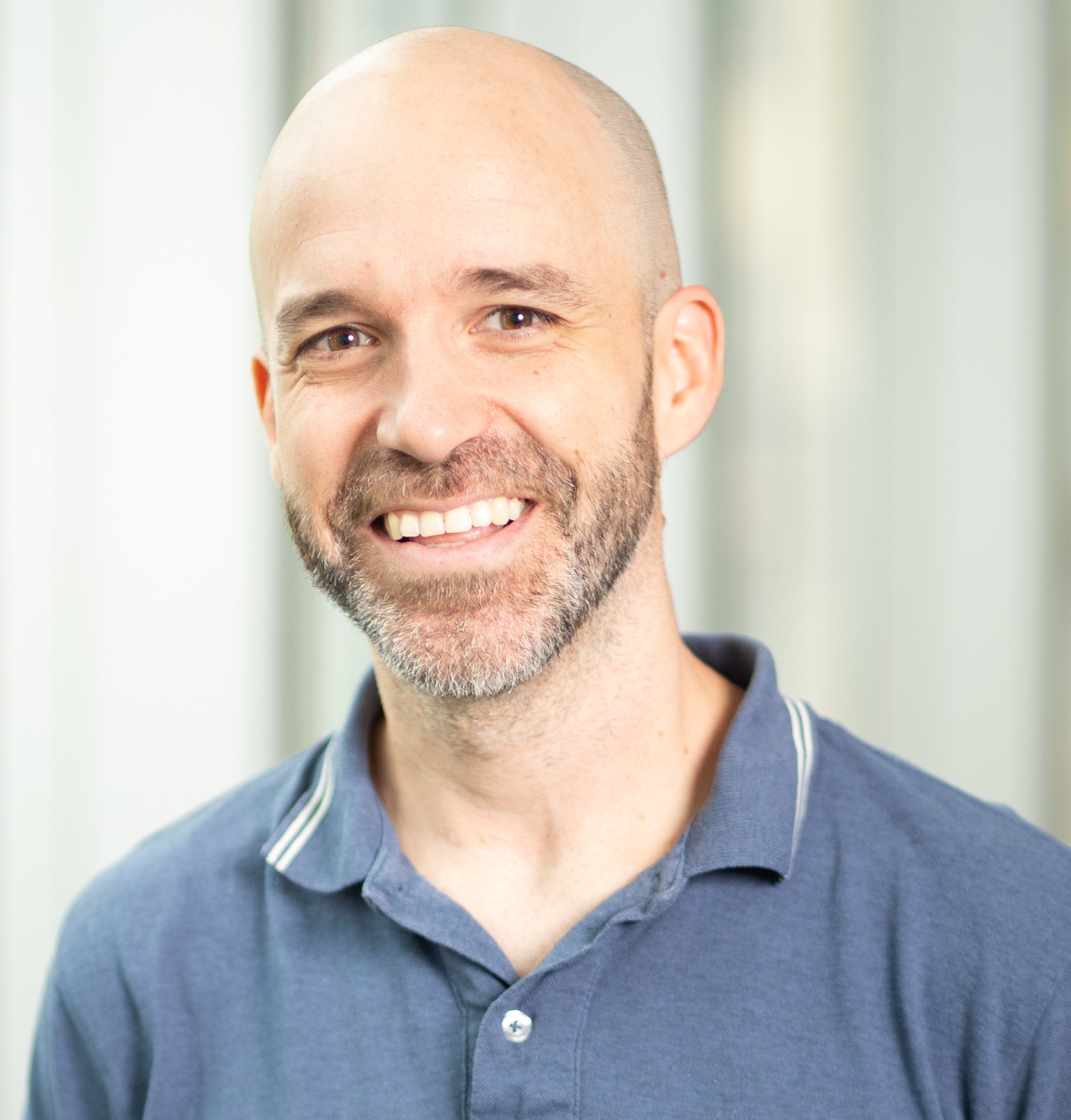}}]{Miguel Catalan-Cid} received the master’s and PhD degrees from the Polytechnic University of Catalonia (UPC), Spain, in 2008 and 2016, respectively. He is currently a Senior Researcher in the Mobile and Wireless Internet Group at i2CAT in Barcelona, Spain. His research interests include 5G/6G mobile networks, software-defined networking, intelligent RAN control, and O-RAN architecture. 
\end{IEEEbiography}
\vspace{-1cm}

\begin{IEEEbiography}[{\includegraphics[width=1in,height=1.25in,clip,keepaspectratio]{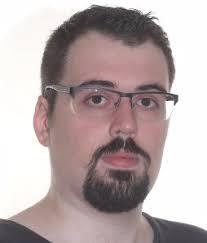}}]{Joan Josep Aleixendri} received the bachelor’s
degree in computer science from the Universitat Politècnica de Catalunya in 2016. He is a Senior R\&D Engineer for the Mobile
and Wireless Internet Group of i2CAT in Barcelona, Spain. His
research topics are software-defined networks, radio access control and management systems, O-RAN technologies and mobile networks.
\end{IEEEbiography}
\vspace{-1cm}

\begin{IEEEbiography}[{\includegraphics[width=1in,height=1.25in,clip,keepaspectratio]{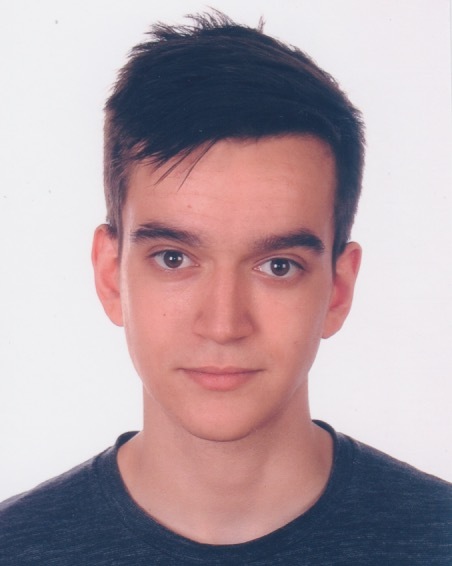}}]{Jorge Pueyo} is a PhD student in Computer Vision at the Polytechnic University of Catalonia (UPC). Master in Advanced Telecommunication Technologies with Deep Learning Specialization by the UPC. Currently doing research in the field of Computer Vision, especially applied to 3D content. Previously part of the  Mobile and Wireless Internet Group at i2CAT research center. 
\end{IEEEbiography}
\vspace{-1cm}

\begin{IEEEbiography}[{\includegraphics[width=1in,height=1.25in,clip,keepaspectratio]{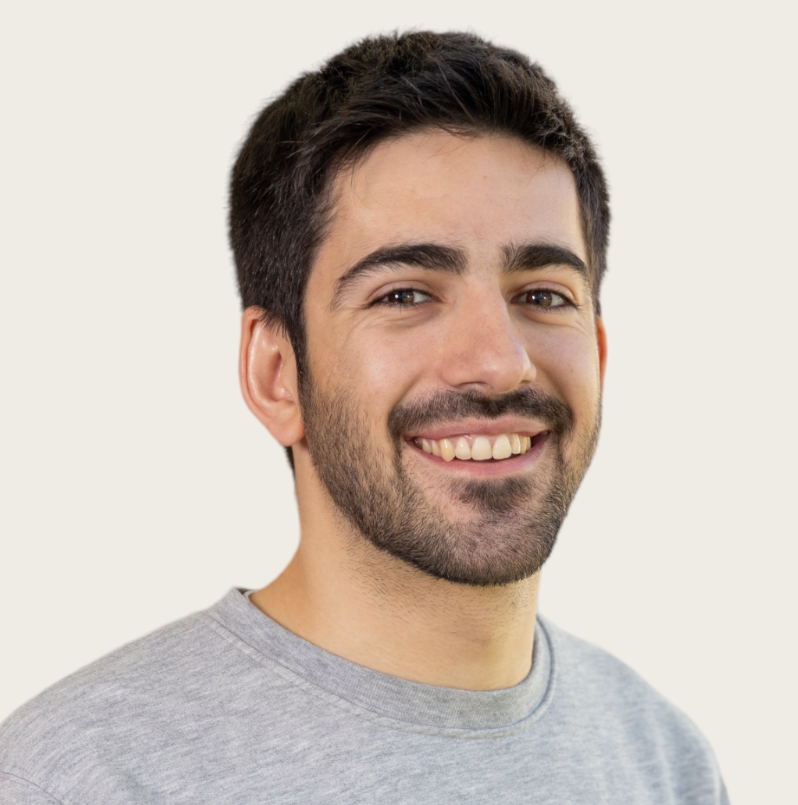}}]{Pau Tomas} received the B.Sc. and M.Sc. degrees in Telecommunications Engineering from the Polytechnic University of Catalonia (UPC), Barcelona, Spain. He is currently a 5G RAN Integration Engineer working on the integration and validation of 5G terrestrial and non-terrestrial networks, edge computing, and network automation. Previously part of the  Mobile and Wireless Internet Group at i2CAT research center.
\end{IEEEbiography}
\vspace{-1cm}

\begin{IEEEbiography}[{\includegraphics[width=1in,height=1.25in,clip,keepaspectratio]{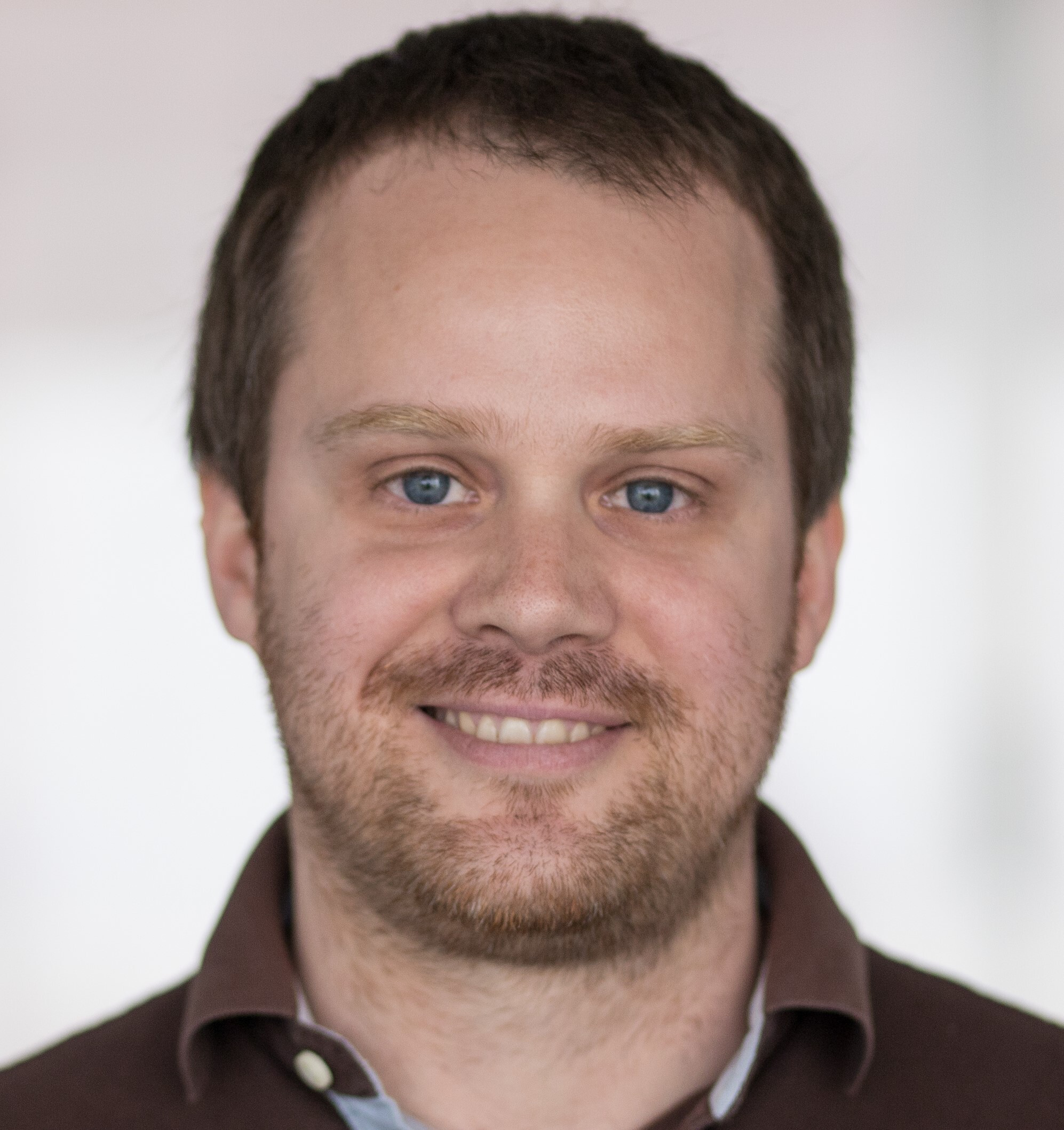}}]{Daniel Camps-Mur} received the master’s and Ph.D. degrees from the Polytechnic University of Catalonia in 2004 and 2012, respectively. He is currently Director of Technology Strategy and leads the Mobile and Wireless Internet Group at i2CAT in Barcelona, Spain. Previously, he was a Senior Researcher with NEC Network Laboratories, Heidelberg, Germany.
\end{IEEEbiography}

%\input{appendix}

%\begin{IEEEbiographynophoto}{Ignacio Ramos}
%(S'12) received the B.S. degree in electrical engineering from the University of Illinois at Chicago in 2009, and is currently working toward the Ph.D. degree at the University of Colorado at Boulder. From 2009 to 2011, he was with the Power and Electronic Systems Department at Raytheon IDS, Sudbury, MA. His research interests include high-efficiency microwave power amplifiers, microwave DC/DC converters, radar systems, and wireless power transmission.

%% insert where needed to balance the two columns on the last page with
%%\newpage


% ==== SWITCH OFF the BIO for submission
% ==== SWITCH OFF the BIO for submission


\vfill


\end{document}